\documentclass[12pt]{article}
\input{epsf}
\usepackage{psfig,epsf}
\usepackage{epsf,cite,epsfig,psfig,float,amssymb,stmaryrd,latexsym}
\usepackage{graphics,psfrag}
\usepackage{a4wide,graphicx}
\usepackage{cite}
\usepackage{amsmath}

\newcommand{\be}{\begin{equation}}
\newcommand{\ee}{\end{equation}}
\newcommand{\ba}{\begin{eqnarray}}
\newcommand{\ea}{\end{eqnarray}}

\newcommand{\ds}{\displaystyle}

\newcommand{\qvec}{\mbox{\boldmath $q$}}
\newcommand{\kvec}{\mbox{\boldmath $k$}}
\newcommand{\lvec}{\mbox{\boldmath $l$}}
\newcommand{\pvec}{\mbox{\boldmath $p$}}
\newcommand{\Pcapvec}{\mbox{\boldmath $P$}}

\newcommand{\Svec}{\mbox{\boldmath $S$}}

\newcommand{\sigmavec}{\mbox{\boldmath $\sigma$}}
\newcommand{\tauvec}{\mbox{\boldmath $\tau$}}
\newcommand{\pivec}{\mbox{\boldmath $\pi$}}

\newcommand{\Sigmacapvec}{\mbox{\boldmath $\Sigma$}}

\begin{document}

\title{
Two pion mediated scalar isoscalar NN interaction in the nuclear medium}

\author{Murat M. Kaskulov, E. Oset and M.J. Vicente Vacas\\
{\small Departamento de F\'{\i}sica Te\'orica and IFIC,
Centro Mixto Universidad de Valencia-CSIC,} \\
{\small Institutos de
Investigaci\'on de Paterna, Aptd. 22085, 46071 Valencia, Spain}\\
}

\date{\today}

\maketitle 

\begin{abstract} 
We study the modification of the nucleon nucleon interaction in a nuclear
medium in the scalar isoscalar channel, mediated by the exchange of two
correlated ($\sigma$ channel) or uncorrelated pions. For this purpose
we use a standard approach for the renormalization of pions in nuclei. 
The corrections obtained for the $NN$ interaction in the medium in this
channel are of the order of 20$\%$ of the free one in average, and the 
consideration
of short range correlations plays an important role in providing these
moderate changes. Yet, the corrections are sizable enough to suggest further
studies of the stability and properties of nuclear matter.
\end{abstract}

\newpage
\section{Introduction}
\small
 
 The determination of the binding energy of nuclei starting from 
 realistic $NN$ potentials is one of the subjects which has received 
 permanent attention from the early days when the Brueckner-Bethe-Goldstone 
 (BBG) equation introduced methods to overcome the strong repulsion of the 
 nuclear forces at short distances.
 
 At present, several many body techniques compete to accurately determine the 
 binding energy of nuclear matter starting from the realistic NN potentials. 
 One of them is the correlated basis functions (CBF) \cite{CBF,fafa}, 
 which follows the line of the Hypernetted Chain Approach (FHNC) 
 \cite{FHNC/SOC}. Another one follows the traditional BBG 
 approach \cite{BBG}, and, although costly numerically, the variational 
 Monte Carlo method (MC) has allowed to make, in principle, exact 
 calculations, although limited to nuclei with small or medium value of 
 A \cite{VMC}. Methods like the selfconsistent treatment of the nucleon 
 selfenergy have also introduced new advances in the field 
 \cite{rpd89,Muther:2000qx,Dickhoff:2004xx}.
  
  The need for three body forces has also been emphasized and the 
  present status is that it is difficult to be quantitative on the strength
  of these forces, and usually they are parameterized in order to adjust 
  the precise value of the binding energy \cite{VMC,Pieper:2001ap,light}. 
  It has also been noted in \cite{Carlson,Wiringa:1988tp,Dewulf:2003nj}
  that short range correlations play an important 
  role when considering these three body forces.
  
  A common feature of these approaches is that they start from the realistic 
  nucleon nucleon interaction, obtained from fits to NN data and deuteron data.
  They use hence the free NN interaction as input. One of the important 
  ingredients of this interaction is the one pion exchange (OPE).  However, 
  from detailed studies of the pion nuclear interaction  
  it is well known that the pion properties in the nuclear medium are sizably 
  renormalized \cite{Oset:1981ih,Gibbs:1987fd,Ericson:1988gk,Nieves:1991ye,
  Chen:1993nv,Nuseirat:1998is,Lee:2002eq}. 
   
  There is also the question of the intermediate range attraction, which is 
  basic in the binding of nuclei. Models for this interaction would contain 
  $\sigma$ exchange, uncorrelated two pion exchange and omega exchange 
  \cite{Machleidt:1987hj}. In as 
  much as the pion properties are changed in the medium, so should the 
  two pion exchange be modified. Medium effects in the two-pion exchange
  have been investigated in early works like in Ref.~\cite{Muther:1980cq}
  restricting themselves in this case to a subset of two pion exchange diagrams
  with no $\Delta$-isobar intermediate states, by including Pauli blocking
  in the intermediate nucleons. 

The medium modification of the two pion exchange got a new impulse after the
models of the $\pi \pi$ interaction in the medium showed large modifications
\cite{Schuck:1988jn,Aouissat:1994sx}, later on softened by the introduction of
chiral constraints~\cite{Rapp:1995ir,Chiang:1997di}.
The implementation of this medium modified $\pi \pi$ interaction in the
correlated two pion exchange $NN$ potential increased appreciably the $NN$ 
attraction in nuclear matter. This was partially reduced by the consideration
of the chiral constraints in Refs.~\cite{Rapp:1996mi,Rapp:1997ii}. 
In these latter
references the importance of short range correlations which modify the $\pi$
nucleus selfenergy was already discussed. The further use of medium modified
vector meson masses led to improvements in the nuclear matter saturation curve.

A new perspective into this problem has been made possible by studies of meson
meson interaction within chiral unitary 
approaches~\cite{Dobado:1996ps,Oller:1997ti,Kaiser:1998fi,Oller:1998hw,
  Oller:1998zr} which allow to improve the description of the correlated two
pion exchange $NN$ interaction~\cite{Oset:2000gn}, as an alternative to the 
conventional $\sigma$-meson exchange interaction.
This picture of the 
  $\sigma$ exchange was mandatory after extensive studies showing that the
  $\sigma$ is not a genuine resonance, made up of $q \bar{q}$ but just the 
  manifestation of a pion - pion resonance state created by the 
  interaction of the pions, what is 
  called a dynamically generated resonance. This shows up naturally within 
  the context of chiral unitary approaches which use the input 
  of the chiral Lagrangians for the meson meson interaction and extends 
  chiral perturbation theory to implement exactly unitarity in coupled 
  channels \cite{Dobado:1996ps,Oller:1997ti,Kaiser:1998fi,Oller:1998hw,
  Oller:1998zr}. This means the $\sigma$ exchange inside a nuclear medium 
  will also be modified as direct consequence of the change of the pion 
  properties.

 Our aim in the present paper is to start from this new 
 picture for the $\sigma$
 exchange, use also the standard approach  for the uncorrelated two pion
 exchange and modify these in the nuclear medium to see what changes one finds
 from these sources.  
 Further improvements come from the consideration of short range correlations
 not only in the pion selfenergy but also in the vertex functions appearing 
 in the model.
 The changes obtained are moderate, thanks to the
 simultaneous consideration of these nuclear short range effects in the
 calculation. In the absence of these, the renormalization of the $NN$
 interaction is huge. Yet, even the moderate results obtained are large enough
 to motivate further calculations of the nuclear binding and other properties
 of matter.

The paper is organized as follows. In Sec.~II we provide those elements of the
 chiral Lagrangian which are relevant for the present calculations and briefly
 discuss peculiarities of the finite baryon density. In Sec.~III we consider
 the modification of the one pion exchange $NN$ force in the nuclear medium
 and in Sec.~IV we discuss the propagation of two pions in the nuclear matter.
 Sec.~V and~VI are devoted to the in-medium  two pion exchange 
 in the scalar-isoscalar channel, both, correlated and uncorrelated.
 The technical details are relegated to the Appendix.
 
\section{Effective Lagrangian}
 In this section we will briefly specify those elements of the  
 effective  chiral Lagrangian in the meson-baryon sector which are 
 relevant for the subsequent calculations. 

 The effective chiral Lagrangian is written as the sum of a purely mesonic
 Lagrangian $\mathcal{L}_{{M}}$
 and the baryonic Lagrangian $\mathcal{L}_{{B}}$
\begin{equation}
\mathcal{L}_{{eff}} = \mathcal{L}_{M} + \mathcal{L}_{B}
\end{equation}
 Both are organized in a derivative and quark mass expansion.
 The lowest order mesonic Lagrangian $\mathcal{L}_2$ is given by
\begin{equation}
\label{Lpipi}
\mathcal{L}_2 = \frac{f_\pi^2}{4}
\langle \partial^{\mu}U^{\dagger} \partial_{\mu}U +
\chi U^{\dagger} + U \chi^{\dagger} \rangle
    \end{equation}
 and contains the most general low-energy interactions of the
 pseudo-scalar meson octet. In Eq.~(\ref{Lpipi}) the symbol
 $\langle ... \rangle$ indicates the trace in flavor space,
 the Goldstone fields are collected in a unitary
 matrix $U$, $f_{\pi}\simeq 93$~MeV is the pseudoscalar decay constant
 and the leading symmetry-breaking term $\chi$ is linear in the quark masses.
 For $SU(2)$ and in the isospin limit 
 $\chi = \mbox{diag}(m_{\pi}^2,m_{\pi}^2)$.
 The lowest order baryon octet - meson octet Lagrangian reads
\begin{equation}
\label{BarLagr}
\mathcal{L}_{B}^{(1)} = 
\langle \bar B (i \gamma^{\mu} D_{\mu} - m_B ) B \rangle 
+ \frac{D}{2} 
\langle \bar B \gamma^{\mu} \gamma_5  \{u_{\mu},B\}\rangle 
+ \frac{F}{2} 
\langle \bar B \gamma^{\mu} \gamma_5   [u_{\mu},B] \rangle 
\end{equation}
where  
the brackets $[...]$ and
$\{...\}$ denote commutators and anti-commutators, respectively. The
covariant derivative of the $SU(3)$ baryon matrix $B$ is defined as
\begin{equation}
\label{CovarDeriv}
D_{\mu} B = \partial_{\mu} B + [\Gamma_{\mu},B]
\end{equation}
In the absence of external field Eqs.~(\ref{BarLagr}) and (\ref{CovarDeriv})
involve other quantities  
\begin{equation}
u = \sqrt{U},~~~ u^{\mu} =  i u^{\dagger}
\nabla^{\mu} U u^{\dagger},~~~ 
\Gamma_{\mu} = \frac{1}{2}(u^{\dagger} \partial_{\mu} u + u \partial_{\mu} 
u^{\dagger})
\end{equation} 
The $SU(3)$ axial vector coupling constants are determined by neutron
and hyperon $\beta$-decay. One finds $F\simeq 0.51$, $D \simeq
0.76$ and the axial coupling constant is $g_A = F + D \simeq 1.27$. 
In the $SU(2)$ limit the Lagrangian simplifies to
\begin{equation}
\label{ChiLagrangian}
\mathcal{L}_N^{(1)} = \bar{\psi}(i \gamma^{\mu} 
D_{\mu}-m_N) \psi + \frac{D+F}{2} \bar{\psi}
\gamma^{\mu} \gamma_5 {u}_{\mu}  \psi
\end{equation}
where $\psi$ is a two component Dirac field $\psi=(p,n)^{T}$.

%------------------------Figure: Sigma PiPi--------------------------
\begin{figure}[t]
\begin{center}
\includegraphics[clip=true,width=0.99\columnwidth,angle=0.]
{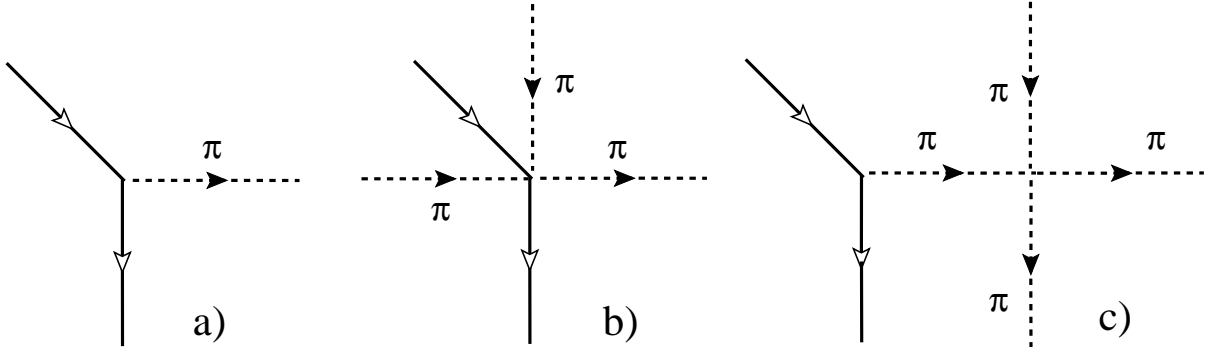}
\caption{\label{3PiVertex} \footnotesize  
The $\pi NN$ vertex (a), the contact $\pi\pi\pi NN$ (b) 
and pion pole (c) terms.}
\end{center}
\end{figure}
%------------------------Figure: Sigma PiPi--------------------------

In the pion-nucleon sector the chiral Lagrangian~(\ref{ChiLagrangian}) 
constrains all possible interactions 
of the pion fields
with fermions at the lowest chiral order we are working. 
For instance, using the exponential parameterization of the 
unitary matrix $U$
\begin{equation}
\label{ExpParameter}
U = \exp \left[ i \frac{\Phi}{f_{\pi}} \right],~~~~ \Phi = \tauvec \pivec
\end{equation}
where $\tauvec$ are the Pauli operators 
and expanding $u$ and $u^{\dagger}$ we obtain the  
pion-nucleon couplings, 
including up to three pion fields
\begin{equation}
\label{uEXPAND}
u_{\mu} = - \frac{\partial_{\mu} \Phi}{f_{\pi}}
+ \frac{1}{24 f_{\pi}^3}(\partial_{\mu} \Phi \cdot \Phi^2
- 2\Phi \partial_{\mu} \Phi \cdot \Phi
+ \Phi^2 \partial_{\mu} \Phi) + \mathcal{O}(\Phi^5)
\end{equation}
If we supplement Eq.~(\ref{uEXPAND}) with the lowest order  
interaction of pions 
as provided by Eq.~(\ref{Lpipi})
\begin{equation}
\mathcal{L}_2 = \frac{1}{48 f_{\pi}^2} 
\langle (\partial_{\mu} \Phi \cdot \Phi
- \Phi \partial^{\mu} \Phi)^2 + m^2_{\pi} \Phi^4 \rangle
\end{equation}
we arrive to the set of Feynman graphs shown in Fig.~\ref{3PiVertex}. Here,
in addition to the standard $\pi NN$ vertex, Fig.~\ref{3PiVertex}a, 
the chiral perturbation theory generates
the contact term of Fig.~\ref{3PiVertex}b (see Appendix A) and the pion 
pole term of Fig.~\ref{3PiVertex}c. These last two 
diagrams are the basic element
in the description of  $\pi N \to \pi \pi N$ reaction near 
the threshold~\cite{Oset:1985wt,Bernard:1995gx,Jensen:1997em,Jaekel:1990es}.
The appearance of the pole terms and contact $3\pi NN$ interactions at the 
same order of the chiral  expansion is crucial for the in-medium calculations
where due to the partial cancellations the physical amplitudes become
independent of the parameterization of $U$ matrix
even in the presence of the nuclear background 
in accord with the equivalence theorem~\cite{Kamefuchi:sb}. One can see
explicitely, that the contact term (b) cancels exactly the off shell part of
the pion pole term (c) coming from the $(q^2-m_{\pi}^2)$ part of the
$\pi \pi \to \pi \pi$ vertex~\cite{Hyodo:2003jw}. 
We shall also see that the off shell part of the
$\pi \pi \to \pi \pi$ amplitude cancel exactly with other terms when we
perform the calculation of the $NN$ interaction in the nuclear medium.

%------------------------Figure:OPE-Graphs-----------------------%
\begin{figure}[t]
\begin{center}
\includegraphics[clip=true,width=0.99\columnwidth,angle=0.]
{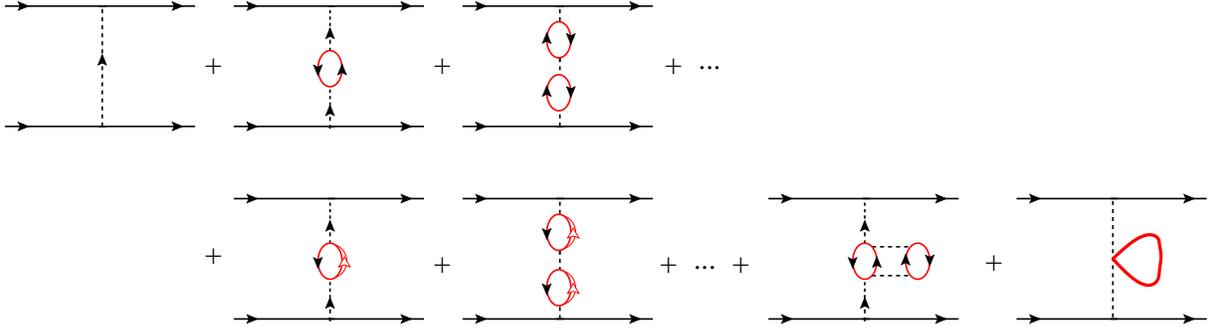}
\caption{\label{OPEPMedimGraphs} \footnotesize 
The OPEP diagrams. The first graph is
the vacuum contribution. The second and third diagrams of the upper 
and first two diagrams of the lower sets of diagrams correspond to the
 $p-h$ and $\Delta-h$ RPA
series. The last two diagrams account for the excitation 
of $2p-2h$ states and the 
contribution of the $S$-wave optical potential, respectively. The latter
diagrams play a minor role and will not be further discussed.}
\end{center}
\end{figure}
%-----------------------%------------------------%-------------------%

\section{One-pion exchange at finite density}
We start with the
one-pion exchange $NN$ potential (OPEP). 
The typical diagrams modifying it 
are shown in Fig.~\ref{OPEPMedimGraphs} where the propagation of 
exchanged pions is distorted by interactions with 
nucleons forming the Fermi sea.  The fermionic bubbles 
describe the decay of the  pion in $p-h$ and $\Delta-h$ states and take 
into account the conventional 
nuclear matter polarization effects. All these diagrams are
responsible for the interaction of two  
nucleons %with initial four momenta $P_1$ and $P_2$ 
in the particle-particle ladder channel with the 
in-medium virtual excitations. In Fig.~\ref{OPEPMedimGraphs}
the first graph is the vacuum contribution. The second and third 
diagrams correspond to the $p-h$ RPA
series and the last diagrams accounts for the excitation of $2p-2h$ states and
the contribution of the $S$-wave optical potential, respectively.
The $p$-wave pion self energy is given by 
\begin{equation}
\label{PolOperator}
\Pi(k,\rho) = \left(\frac{D+F}{2 f_{\pi}} \right)^2 \kvec^2 
\mathcal{U}(k,\rho) 
\left[ 1 -  \left(\frac{D+F}{2 f_{\pi}}\right)^2 g' \mathcal{U}(k,\rho) \right]^{-1}
\end{equation}
where $g'=0.6$ is the Landau-Migdal parameter~\cite{Migdal:1978az}, 
$\rho$ is
the nuclear matter density and
$\mathcal{U}(k,\rho)=\mathcal{U}^d(k,\rho) + \mathcal{U}^c(k,\rho)$
is the Lindhard function  accounting for the direct and crossed 
contributions of $p-h$ and $\Delta-h$ excitations 
with the normalization of the appendix of Ref.~\cite{Oset:1989ey}.

%------------------------Figure:OPE-ForceGraph -----------------------%
\begin{figure}[t]
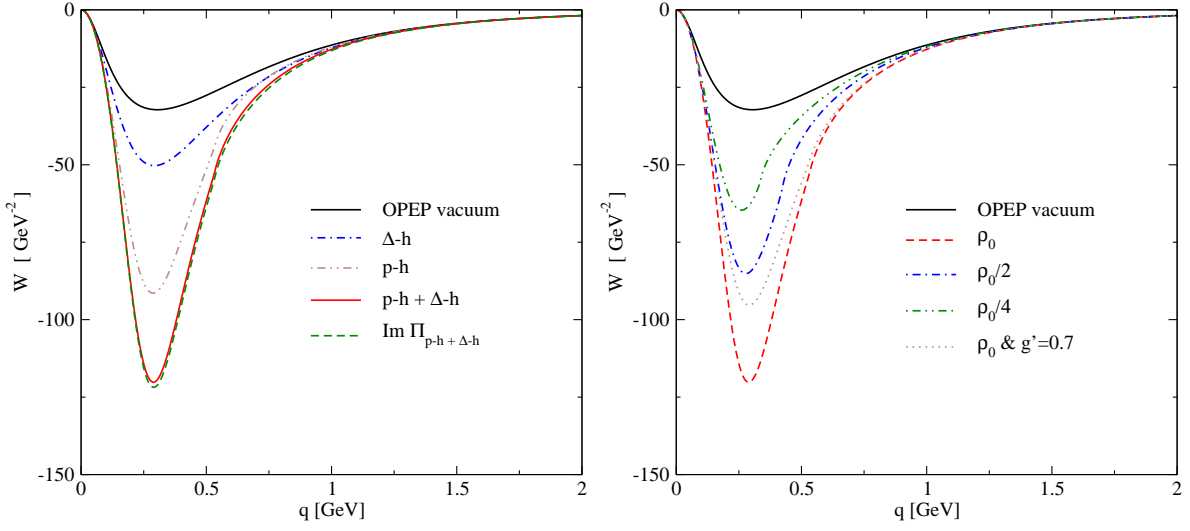

\begin{center}
\includegraphics[clip=true,width=0.48\columnwidth,angle=0.]
{Figure3a.eps}
\includegraphics[clip=true,width=0.48\columnwidth,angle=0.]
{Figure3b.eps}
\caption{\label{OPEPMedium} \footnotesize Left panel: The OPEP in the vacuum 
(solid curve) and at normal $\rho_0=0.16$~fm$^{-3}$ nuclear matter 
density (dashed curve). Right panel: OPEP at several densities.
}
\end{center}
\end{figure}
%-----------------------%------------------------%-------------------%

The OPEP in the momentum space  takes the form
\begin{eqnarray}
\label{OPEPre}
-i V_{OPEP}(\qvec,\rho) %&=& -i \left(\frac{D+F}{2 f_{\pi}}\right)^2 
&=&  - i
W(\qvec,\rho)
\hat{\qvec}_i \hat{\qvec}_j \sigma_{1}^i \sigma_{2}^j \tauvec_1 \tauvec_2
\end{eqnarray}
where we have defined
\begin{equation}
\label{Wopep}
W(\qvec,\rho)=\left(\frac{D+F}{2 f_{\pi}}\right)^2
\qvec^2  F^2(\qvec) \tilde{D}_{\pi}(0,\qvec) 
\end{equation}
where $\tilde{D}_{\pi}(q_0,\qvec)$ is the pion propagator in the medium
\begin{equation}
\label{InMediumGF}
 i\tilde{D}_{\pi}(k) = \frac{i}{k^2-m^2_{\pi} -\Pi(k,\rho) + i 0^+}
\end{equation}
and $F(\qvec)$ stands for a monopole form factor 
$\Lambda^2/(\Lambda^2+\qvec^2)$ with the cut off scale $\Lambda=1$~GeV,
$\qvec$ is a momentum transfer and $\hat{\qvec}=\qvec/|\qvec|$.
Note, that OPEP depends on the real part of the polarization operator only,
since for $q=(0,\qvec)$ one has $\mbox{Im}\Pi(0,\qvec,\rho) =0$.
The well-known vacuum $NN$ amplitude is recovered in Eq.~(\ref{Wopep})
by setting $\Pi=0$  or in the limit $\rho = 0$.

Our results for $W(\qvec,\rho)$ 
are presented in 
Fig.~\ref{OPEPMedium} (left).
The standard vacuum behavior is shown by the solid curve. The dashed curve
represents the modified OPEP at normal nuclear matter density were
we observe an additional strong increase of the strength 
associated with the attractive excitation
of $p-h$ and $\Delta -h$ collective states, with $p-h$ playing a dominant
role. Their individual contributions are shown by 
dot-dot-dashed and dot-dashed curves, respectively.

%------------------------Figure:OPE-ForceGraph -----------------------%
\begin{figure}[t]
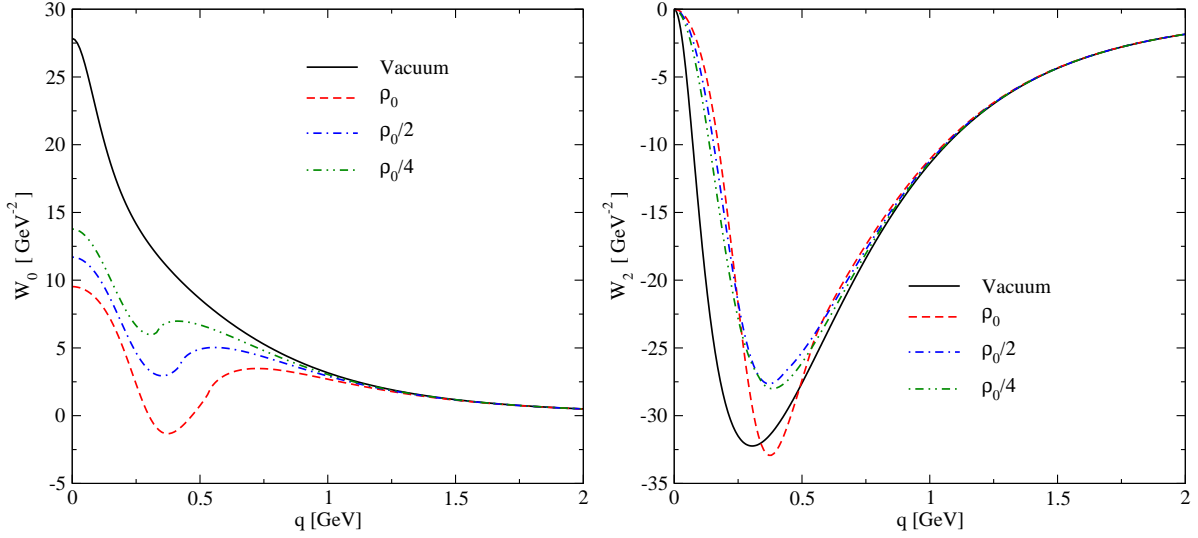

\begin{center}
\includegraphics[clip=true,width=0.48\columnwidth,angle=0.]
{Figure4a.eps}
\includegraphics[clip=true,width=0.486\columnwidth,angle=0.]
{Figure4b.eps}
\caption{\label{OPEPMediumModif} \footnotesize The OPEP with short range
  correlations (solid curves) as a function of the nuclear matter 
density. 
The left and right panels are the central and tensor parts, respectively.}
\end{center}
\end{figure}
%-----------------------%------------------------%-------------------%

The analytic properties of the in-medium pion propagator which we use here  
may be verified by using the dispersion representation 
for the Green function
\begin{eqnarray}
\label{LehmanRepres}
i \tilde{D}_{\pi}(k_0,\kvec)= -i \frac{2}{\pi}  \int_0^{\infty} d x \, x  \,
 \frac{\mbox{Im}  \tilde{D}_{\pi}(x,\kvec)}{k_0^2-x^2 + i 0^+
 }
\end{eqnarray}
In this case the OPEP of Eq.~(\ref{Wopep})  can be written in terms of
the absorptive part of the pion propagator only
\begin{eqnarray}
W(\qvec,\rho) = \frac{2}{\pi} \left(\frac{D+F}{2 f_{\pi}}\right)^2
\qvec^2  F^2(\qvec) \int \limits_0^{\infty} \frac{dx}{x} %-i0^+}
\Big[\mbox{Im} \tilde{D}_{\pi}(x,\qvec)\Big]
\end{eqnarray}
The causality 
requires that both equations must produce the same result.
Indeed as one can see in Fig.~\ref{OPEPMedium} the curves calculated with 
dispersive
and absorptive parts of the pion propagator are practically 
indistinguishable.
We would like to note that a strong modification of the OPEP at 
finite baryonic density 
observed here
is not new and was predicted long time ago by Migdal~\cite{Migdal:1978az}.
There it was also shown that in-medium modified  OPEP helps to explain 
the unnatural parity states in finite nuclei,
for instance, the shift of $0^-$ state in closed shell nuclei. 

In the right panel of Fig.~\ref{OPEPMedium}
we show our combined plot for
a few densities $\rho_0, \rho_0/2$ and $\rho_0/4$. Here, we also show the
sensitivity of our results to the value of the Landau-Migdal parameter $g'$.
We find that the increase of $g'$ from 0.6 to 0.7 makes the OPEP
softer. 
This fact suggests that the proper treatment of the $NN$
short range correlations is important for understanding 
the in-medium properties of the OPEP. 

At this point we would like to mention that in a realistic calculation one
will have to add strong repulsive forces at short distances. This can be done
in a straightforward way using any of many body schemes discussed in the 
introduction. The correlations of this part of the interaction would 
effectively modulate the $\pi$ exchange interaction, introducing the 
correlation parameter $g'$~\cite{Oset:1979bi}. The denominator in 
Eq.~(\ref{InMediumGF})
takes into account this effect between $p$-wave bubbles in the diagrams 
of Fig.~\ref{OPEPMedimGraphs}, but
not between the external nucleon and the contiguous bubble. To account for
this we make the separation between the longitudinal and transverse parts 
of the pion effective interaction~\cite{Alberico:1981sz,Oset:1987re}
\begin{equation}
\left(\frac{D+F}{2 f_{\pi}}\right)^2 F(\qvec)^2 
\frac{q_i q_j}{q_0^2-\qvec^2-m_{\pi}^2 + i0^+} \longrightarrow
\mathcal{V}_l(q) \, \hat{q}_i  \hat{q}_j 
+ \mathcal{V}_t(q) (\delta_{ij} - \hat{q}_i  \hat{q}_j)
\end{equation}
where $\hat{q}_i$ is the Cartesian component of the unit vector 
$\hat{\qvec}=\qvec/|\qvec|$ and
\begin{equation}
\label{Vl}
\mathcal{V}_l(q) = \left(\frac{D+F}{2 f_{\pi}}\right)^2 
\left[ \frac{\qvec^2}{q_0^2-\qvec^2-m_{\pi}^2 + i 0^+}  + g'\right]F(\qvec)^2 
\end{equation}
\begin{equation}
\label{Vt}
\mathcal{V}_t(q) = \left(\frac{D+F}{2 f_{\pi}}\right)^2 \, g' \, F(\qvec)^2 
\end{equation}
When we perform the sum of diagrams in Fig.~\ref{OPEPMedimGraphs} 
then we get
\begin{eqnarray}
V_{OPEP}(q,\rho) &=&
\left[ \frac{\mathcal{V}_l(q)}{1 - \mathcal{U}(q,\rho) \mathcal{V}_l(q) } 
\hat{q}_i \hat{q}_j + 
\frac{\mathcal{V}_t(q)}{1 - \mathcal{U}(q,\rho) \mathcal{V}_t(q) } 
(\delta_{ij} - \hat{q}_i \hat{q}_j)
\right]\sigma_1^i \sigma^j_2   \tauvec_1 \tauvec_2  \nonumber \\
&=& W_0(q,\rho) \sigmavec_1 \sigmavec_2  \tauvec_1 \tauvec_2 
+
W_2(q,\rho) (\sigmavec_1 \hat{\qvec} 
\sigmavec_2 \hat{\qvec} - 
\frac{1}{3}\sigmavec_1 \sigmavec_2)  \tauvec_1 \tauvec_2 
\end{eqnarray}
\begin{eqnarray}
W_0(q,\rho) &=&
\frac{1}{3} \left[\frac{\mathcal{V}_l(q)}{1-\mathcal{U}(q,\rho)
\mathcal{V}_l(q)}  + 
\frac{2\mathcal{V}_t(q)}{1-\mathcal{U}(q,\rho)\mathcal{V}_t(q)}
     \right] \\
W_2(q,\rho)&=&
\frac{\mathcal{V}_l(q)}{1-\mathcal{U}(q,\rho)\mathcal{V}_l(q)}
-  \frac{\mathcal{V}_t(q)}{1-\mathcal{U}(q,\rho)\mathcal{V}_t(q)}
\end{eqnarray}
where we have explicitely separated the central and the tensor parts of the 
interaction. In Fig.~\ref{OPEPMediumModif} 
we show the results for the central $W_0$ (left panel)
and tensor $W_2$ (right panel) parts (omitting
the spin-isospin operators) as a function of the baryonic density.
Note that in the limit $\rho \to 0$, $W_2 = W$ where $W$ is given by 
Eq.~\ref{Wopep}.

\section{Two pions in the medium}
In this section we turn to the dynamics of the two pion system in the nuclear
medium. Our main interest here is the propagation of 
two $S$-wave pion pairs. As it is well known in the $S$-wave scattering
the use of the proper unitarization schemes
lead to the generation of the $\sigma$-meson.  
Later on we will use this result
for the in-medium $NN$ interaction mediated by exchange of two 
correlated pions
in the  scalar-isoscalar $\sigma$-meson channel.

%------------------------Figure:TriangleGraph -----------------------%
\begin{figure}[b]
\begin{center}
\includegraphics[clip=true,width=0.85\columnwidth,angle=0.]
{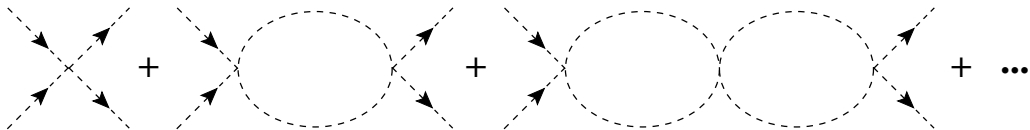}
\caption{\label{UnitSeries} \footnotesize The unitary series representing 
$\pi \pi$ scattering amplitude.}
\end{center}
\end{figure}
%------------------------%------------------------%-------------------%

For the $\pi_a  \pi_b \to \pi_c  \pi_d$
scattering process, defined by the Cartesian isospin indices $a,...$,
the use of the standard $\chi$PT procedure in expanding the
$\mathcal{L}_{2}$ of Eq.~(\ref{Lpipi}) to  order $\mathcal{O}(\pivec^4)$
results in the tree level contact interaction
\begin{eqnarray}
\label{TreePiPi}
-i V_{\pi \pi}^{ab \to cd} =
\delta_{ab} \delta_{cd}
A(s)
+ \delta_{ac} \delta_{bd}
A(t)
+ \delta_{ad} \delta_{bc}
A(u),
\end{eqnarray}
where
\begin{equation}
A(s) = \frac{i}{f_{\pi}^2} \Big( s- M_{\pi}^2 - \frac{1}{3}
  \sum_{i=a,b,c,d} \Lambda_i \Big) + \mathcal{O}(q^4),
\end{equation}
and $\Lambda_{i} = k_i^2 - M_{\pi}^2$ is the off-shell part of
the invariant $\pi \pi$ amplitude. 
At this order of the pion field expansion the isoscalar $S$-wave $\pi \pi$
 partial amplitude ($L=0$) is obtained from the standard decomposition
\begin{equation}
V_{\pi \pi}^{L,I=0} = \frac{1}{2} \frac{1}{(\sqrt{2})^{\alpha}}
\int_{-1}^{1} d \cos\theta~ {P}_L(\cos\theta)~ V_{\pi
\pi}^{I=0}(\theta)
\end{equation}
where ${P}_L(\cos\theta)$ are the Legendre polynomials and
$(\sqrt{2})^{\alpha}$ accounts for the
statistical factor occurring in states with identical particles:
$\alpha =2$ for $\pi \pi \to \pi \pi$ in the unitary normalization of the 
states~\cite{Oller:1997ti}. The tree level scalar-isoscalar
$\pi \pi$ scattering amplitude  is
\begin{equation}
\label{VPiPi00}
V_{\pi \pi}^{L=I=0} = - \frac{1}{f_{\pi}^2}
\Big(s - \frac{M_{\pi}^2}{2} - \frac{1}{3} \sum_i \Lambda_i \Big).
\end{equation}
In Eq.~(\ref{VPiPi00}) the off shell part depends on  choice of $U$ 
and is equal to zero for
on mass shell pions. 

Following Ref.~\cite{Oller:1997ti} and using the Bethe-Salpeter equation 
we unitarize
the $S$-wave $\pi \pi$ scattering amplitude (see Fig.~\ref{UnitSeries}) 
\begin{equation}
\label{TPiPiUnitary}
V_{\pi \pi}^{L=I=0} \to {T}^{L=I=0}_{\pi \pi} = 
\left[
{\ds - {f^2_{\pi}}\left({ s - \frac{M^2_{\pi}}{2}}\right)^{-1} -
  \ds  G_{\pi \pi}(s)} \right]^{-1},
\end{equation}
where $G_{\pi \pi}(s)$ is a scalar two-pion loop function
\begin{equation}
\label{GPiPi}
G_{\pi\pi}(s) = 
i \int \frac{d^4 k}{(2 \pi)^4} \frac{1}{[(P-k)^2-M_{\pi}^2+i0^+] 
(k^2-M_{\pi}^2+i0^+)}
\end{equation}
where $P^{\mu} = (P_0,\Pcapvec)$ and $s=P^2$.  
The $G_{\pi\pi}(s)$  function is analytic with a 
cut along the positive real axis starting at the $\pi \pi$ threshold. 
Note that, Eq.~(\ref{TPiPiUnitary}) contains a pole 
in the second Riemann
sheet corresponding to the $\sigma$-meson with mass and width 
$M - i \Gamma/2 \simeq 450 - i 221$~MeV.

%------------------------Figure:TriangleGraph -----------------------%
\begin{figure}[t]
\begin{center}
\includegraphics[clip=true,width=0.97\columnwidth,angle=0.]
{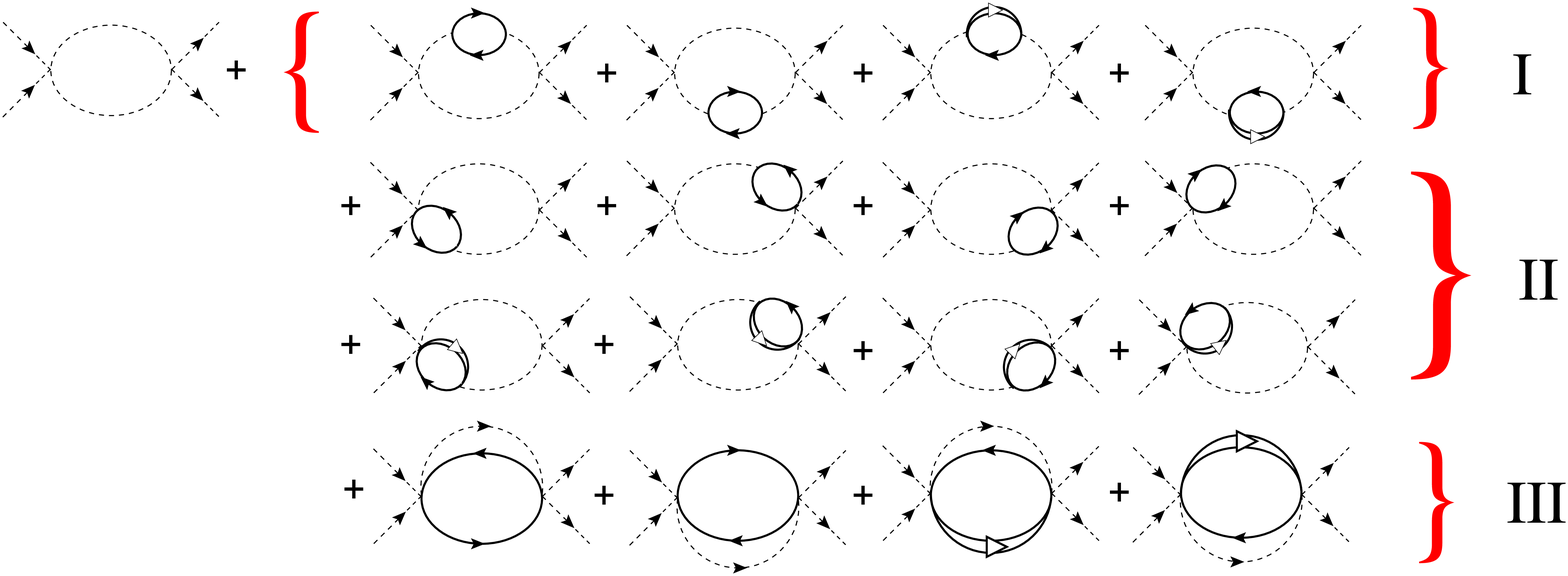}
\caption{\label{RenormPiPiInMedium} \footnotesize 
Renormalization of the in-medium
$\pi \pi$ amplitude including $p$-hole and $\Delta$-hole
excitations.}
\end{center}
\end{figure}
%------------------------%------------------------%-------------------%

In the nuclear medium the $S$-wave $\pi \pi$ scattering amplitude and 
therefore 
the $\sigma$-meson get renormalized and explicit calculations were done in
Refs.~\cite{Chanfray:1999nn,Chiang:1997di}.
The diagrams at one loop level are shown in 
Fig.~\ref{RenormPiPiInMedium}. For instance, the amplitudes corresponding
to the insertion of fermion bubbles in the upper meson line are given by 
\begin{eqnarray}
V_{\mbox{\tiny I}}^{up}(s) &=& i
\left( \frac{D+F}{2 f_{\pi}} \right)^2
\int \frac{d^4 k}{(2 \pi)^4} \kvec^2 \mathcal{U}(k) D^2_{\pi}(k) D_{\pi}(P-k)
\Big[ V_{on}(s) + \frac{1}{3 f_{\pi}^2} \sum_i \Lambda_i \Big]^2 \\
V_{\mbox{\tiny II}}^{up}(s) &=&  \frac{-i2}{3f_{\pi}^2}
\left( \frac{D+F}{2 f_{\pi}} \right)^2
\int \frac{d^4 k}{(2 \pi)^4} \kvec^2 \mathcal{U}(k) D_{\pi}(k) D_{\pi}(P-k)
\Big[ V_{on}(s) + \frac{1}{3 f_{\pi}^2} \sum_i \Lambda_i \Big] \\
V_{ \mbox{\tiny III}}^{up}(s) &=& \frac{i}{9f_{\pi}^4}
\left( \frac{D+F}{2 f_{\pi}} \right)^2
\int \frac{d^4 k}{(2 \pi)^4}  \kvec^2 \mathcal{U}(k) D_{\pi}(P-k)
\end{eqnarray}
As was shown in Refs.~\cite{Chanfray:1999nn,Chiang:1997di},
in the center of mass frame of the 
two pions the off shell part ($\Lambda$ terms) in
$V_{I}^{up}$ cancels exactly the $V_{II}^{up}$ and $V_{III}^{up}$ terms.
And one is left  only with the diagrams of type (I) but 
with the on shell $\pi \pi$ amplitude
\begin{eqnarray}
V^{up}(s) &=& i
\left( \frac{D+F}{2 f_{\pi}} \right)^2
V_{on}^2(s) 
\int \frac{d^4 k}{(2 \pi)^4} \kvec^2 \mathcal{U}(k) 
D^2_{\pi}(k) D_{\pi}(P-k) %\\
\end{eqnarray}
It is straightforward to iterate the $p-h$ and $\Delta-h$ excitations in 
Fig.~\ref{RenormPiPiInMedium} (I) and the loop function, $T(s)$, is given by
\begin{equation}
T(s) = V_{on}^2(s) \cdot \tilde{G}_{\pi\pi}(s)
\end{equation}
where $ \tilde{G}_{\pi\pi}(s)$ is in-medium modified scalar loop integral
\begin{equation}
\label{GPiPimediumI}
\tilde{G}_{\pi\pi}(s) = i \int \frac{d^4 k}{(2 \pi)^4} 
\tilde{D}_{\pi}(k) \tilde{D}_{\pi}(P-k)
\end{equation}
Using the spectral representation for the in-medium pion propagators, 
Eq.~(\ref{InMediumGF}), we get for the loop function
\begin{eqnarray}
\label{GpipiGen}
\tilde{G}_{\pi\pi}(P) = 
\frac{2}{\pi^2} \int \frac{d \kvec}{(2 \pi)^3}
\int  \limits_0^{\infty} dx
\Big[\mbox{Im} \tilde{D}_{\pi}(x,\kvec)\Big]  
 \int  \limits_{x}^{\infty} d y
\Big[\mbox{Im} \tilde{D}_{\pi}(y-x,\Pcapvec-\kvec)\Big] \nonumber \\
\times \frac{y}{(P_0+y)(P_0-y+i0^+)} 
\end{eqnarray}

\begin{figure}[t]
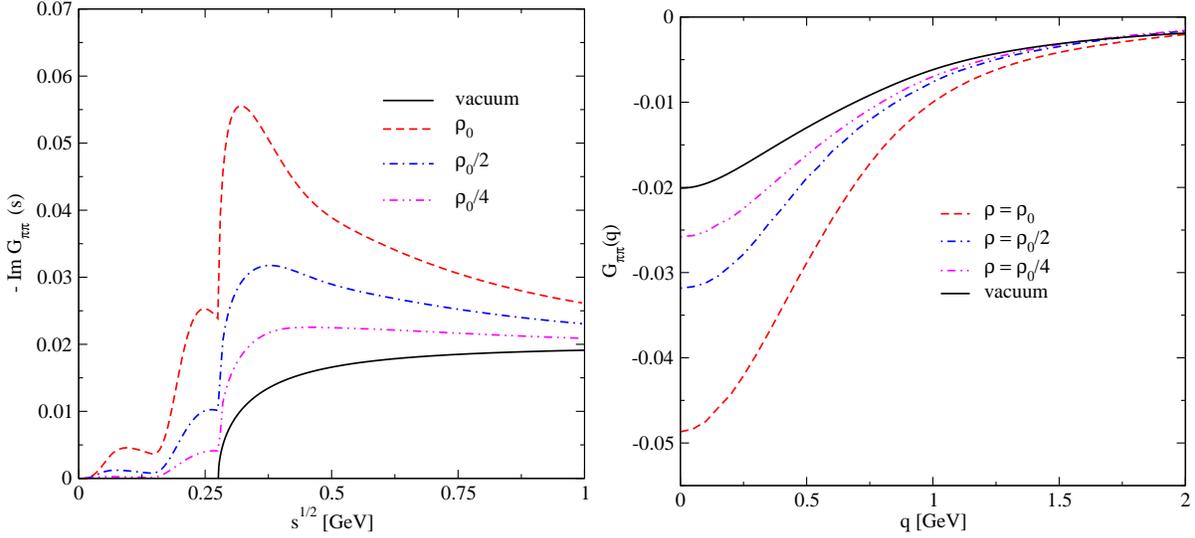

\begin{center}
\includegraphics[clip=true,width=0.48\columnwidth,angle=0.]
{Figure7a.eps}
\includegraphics[clip=true,width=0.486\columnwidth,angle=0.]
{Figure7b.eps}
\caption{\label{ImGpipiMedium} \footnotesize The 
imaginary part of the in-medium scalar loop function $G_{\pi \pi}$ in the $\pi
\pi$ CM frame (left panel) with the normal dressing of the
pion propagators including $p-h$, $\Delta-h$ and $2p-2h$ excitations. 
The right panel shows the loop function
in the space like region $P^{\mu}=(0,\qvec)$.}
\end{center}
\end{figure}

We refer to Ref.~\cite{Cabrera:2005wz} where different aspects 
of $S$-wave $\pi \pi$ scattering in the nuclear medium 
are discussed
and also the behavior of the $\sigma$-meson mass and width
at finite  baryonic density is addressed. But here, we would like to
illustrate the impact of the
nuclear medium on the $\pi \pi$ system. For that
consider the imaginary part of the loop 
function in the $\pi \pi$ 
center of mass frame $P^{\mu}=(P_0,0)$ with $P_0=\sqrt{s}$. This
situation  is
relevant for the in-medium $\pi \pi$ scattering and 
contains the proper information 
about
the dynamics of the pole position of $\sigma$  at finite density. 
Our results for the imaginary part of the scalar loop function  for several 
densities are shown in Fig.~\ref{ImGpipiMedium}(left panel). 
The solid curve correspond to the vacuum loop function
which can be obtained from Eq.~(\ref{GpipiGen}) by substituting 
the imaginary part of 
in-medium pion propagators
by their vacuum expressions 
\begin{equation}
\label{ImPartVacuum}
\mbox{Im} \, G_{\pi\pi}(\sqrt{s}) 
= -\frac{1}{16 \pi}
\sqrt{1 - \frac{4 m_{\pi}^2}{s}}
\end{equation}
which is proportional to the 
two-body phase space of two particles (pions).
As one can see in Fig.~\ref{ImGpipiMedium}~(left) 
the effect of the medium is remarkable due to the increase of
available for pions phase space because of
additional pion decay branches like
$p-h$, $\Delta-h$ and $2p-2h$. 

%------------------------Figure: Sigma PiPi--------------------------
%------------------------------------------------------------------

The kinematics relevant for the $NN$ force, where two nucleons interact
by exchange of mesons is defined
by the moving reference frame 
where $P^{\mu}=(0,\qvec)$. In this case the $\pi \pi$ interaction in the
medium has to be generalized from the results 
in~\cite{Chanfray:1999nn,Chiang:1997di} since there $P_0 \neq 0$ but 
$\qvec=0$. We repeat all the steps that led to cancellations 
in~\cite{Chanfray:1999nn,Chiang:1997di} and find again the same cancellations
as before but with some remnant $\qvec^2$ dependent terms vanishing in 
the limit $\qvec^2 = 0$. To evaluate
these terms we simplify the calculation assuming $\qvec$ relatively small
(this is fine for momenta below the Fermi momentum). Concretely, we assume
$$1)~~~ \qvec^2/k_{max}^2 \ll 1,$$
where $k_{max}$ is the cut off in the three momentum 
(of the order of $1$~GeV in~\cite{Oller:1997ti}) that one uses to regularize
the $G_{\pi \pi}$ function. We also assume
$$2)~~~ \kvec^2 D_{\pi}(k) \simeq -1$$
which implies $k_{max}^2 \gg m_{\pi}^2$ as it is the case.
And 3) we expand $D_{\pi}(k)$ in terms of $D_{\pi}(P-k)$ and vice-versa
to relate different terms. After all this is done we find that the corrections
can be taken into account by means of the change in the Lindhard function
\begin{equation}
\label{Umodif}
\mathcal{U}(k) \to \mathcal{U}(k) \left[ 1 + 
\frac{ \qvec^2}{3}  \left(s - \ds \frac{m_{\pi}^2}{2}\right)^{-1}
+ 
\frac{\qvec^4}{3}  \left(s - \ds \frac{m_{\pi}^2}{2}\right)^{-2}
\right] 
\end{equation}
The expression in brackets in Eq.~(\ref{Umodif}) is $\simeq 1$ for very small $q$ and also
$\simeq 1$ for $\qvec^2 \gg m_{\pi}^2/2$. Hence with very good approximation
we can take the bracket equal to unity and thus there are no other corrections
to be done to the result of~\cite{Chanfray:1999nn,Chiang:1997di} except
the obvious one of changing $s \to - \qvec^2$. For these value of $s$ the 
$G_{\pi \pi}$ is only real, contrary to the case studied 
in~\cite{Chanfray:1999nn,Chiang:1997di}. 
Results for $G_{\pi \pi}(\qvec^2)$ 
can be seen in Fig.~\ref{ImGpipiMedium} (right panel) for different nuclear
densities $\rho_0$, $\rho_0/2$ and $\rho_0/4$. We can see that the 
corrections are sizable particularly at small values of $|\qvec|$.

\section{In-medium renormalization of the correlated 
two pion ($\sigma$-meson) exchange}

\begin{figure}[t]
\begin{center}
\includegraphics[clip=true,width=0.9\columnwidth,angle=0.]
{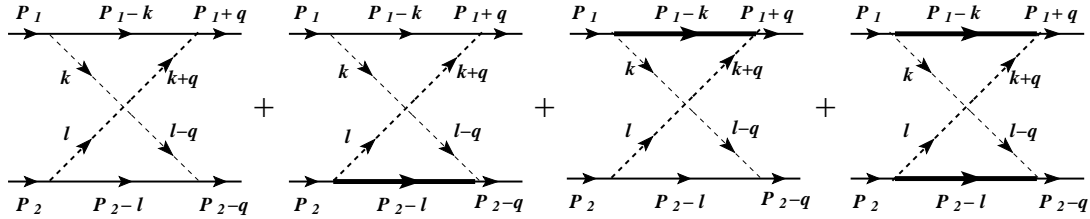}
\caption{\label{CTPEvacuum.eps} \footnotesize Diagrams representing the
  correlated two pion exchange with $NN$, $N\Delta$ and $\Delta \Delta$
  intermediate states.}
\end{center}
\end{figure}

\begin{figure}[h]
\begin{center}
\includegraphics[clip=true,width=0.8\columnwidth,angle=0.]
{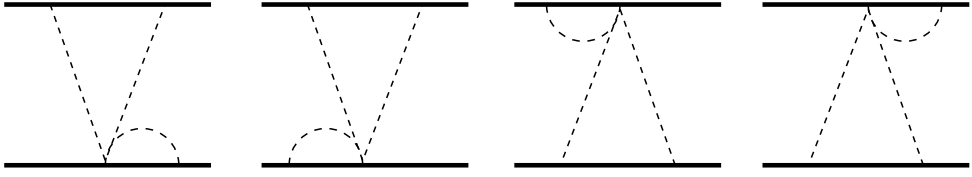}
\caption{\label{SETCans} \footnotesize The set of diagrams which cancel
  the off-mass shell part of the scalar-isoscalar correlated two-pion 
  exchange.}
\end{center}
\end{figure}

The correlated two pion exchange (CrTPE) in the scalar-isoscalar channel, 
the equivalent to a $\sigma$ exchange in
meson exchange models~\cite{Machleidt:1987hj}, or correlated two pion exchange
in the dispersion relations in~\cite{Lin:1990cx} was studied 
in~\cite{Oset:2000gn,Kaiser:1998wa} within the context of chiral Lagrangians.
One starts from the diagrams of Fig.~\ref{CTPEvacuum.eps}, where the $\pi \pi
\to \pi \pi$ scattering shows the  off shell ambiguities. 
To avoid these ambiguities with isoscalar exchange 
it was stated in Ref.~\cite{Kaiser:1998wa}
that one must include the subset of diagrams of Fig.~\ref{SETCans}
to find cancellations
of the off shell $\pi \pi$ isoscalar amplitude. 
This statement was rigorously verified in Ref.~\cite{Oset:2000gn},
where it was shown that the consideration of these subset
of chiral diagrams, Fig.~\ref{SETCans}, 
including the contact $3\pi NN$ interactions, 
results in the cancellation of the off-shell part
of the $\pi \pi$ amplitude, and the on-shell part of the $\pi \pi$ 
amplitude can be
factorized out from the loop integrals.
One step forward was 
given in~\cite{Oset:2000gn}, where iteration of the $\pi \pi$ interaction,
through the Bethe-Salpeter equation, was done by means of which a simple analytical expression was
obtained for the correlated two pion exchange in the scalar-isoscalar channel
\begin{equation}
\label{Vcrtpesigma}
V_{\sigma}(t)= 6 V^2(t) \left[
{\ds {-f^2_{\pi}}\left({ t - \frac{M^2_{\pi}}{2}}\right)^{-1} -
  \ds  G_{\pi \pi}(t)} \right]^{-1}
\end{equation}
where $t = -\qvec^2$ in the $NN$ c.m. frame.
The vertex function $V(t)=V_N(t)+V_{\Delta}(t)$ 
for the triangle loop with two mesons and one
baryon propagator, including $N$ and $\Delta$ intermediate states, 
is evaluated in~\cite{Oset:2000gn}
using a cut off in $|\kvec|$ of about 1~GeV.
Note that, the  bracket in Eq.~(\ref{Vcrtpesigma}) contains a pole
in the $s$-channel corresponding to the $\sigma$-meson. This
restores the relation to
the $\sigma$-meson exchange, which now enters the formalism as a
dynamical resonance in the $\pi \pi$ system~\cite{Oset:2000gn} 
(see also related discussions in Refs.~\cite{Kaskulov:2004kr,Kaskulov:2004vj}).

%------------------------Figure: Sigma PiPi--------------------------
%------------------------------------------------------------------

%------------------------Figure: Sigma PiPi--------------------------
%------------------------------------------------------------------

%------------------------Figure:TriangleGraph -----------------------%
\begin{figure}[t]
\begin{center}
\includegraphics[clip=true,width=0.99\columnwidth,angle=0.]
{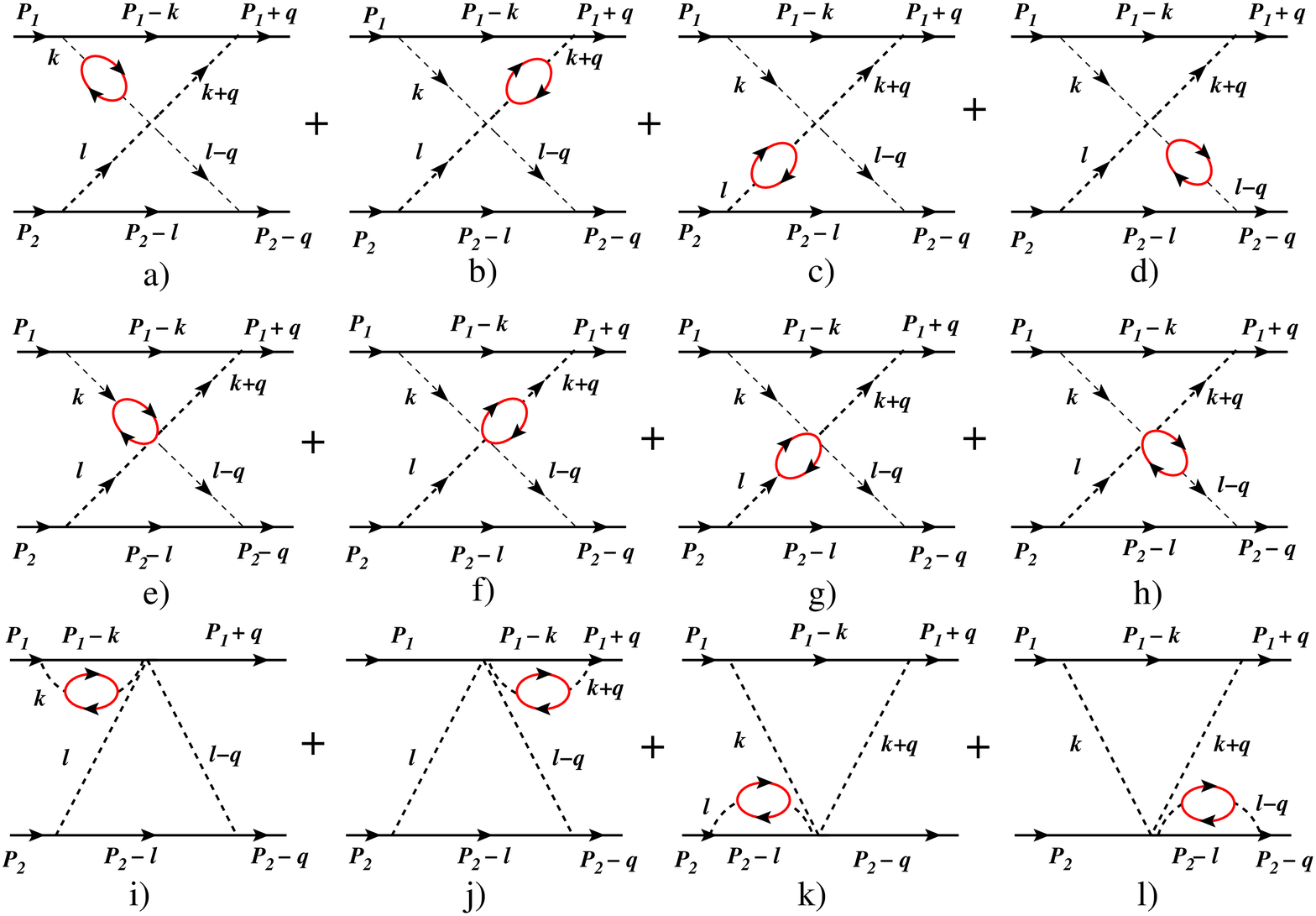}
\caption{\label{CorrTPECrMedium} \footnotesize 
The in-medium diagrams involving the $p$-wave pion dressing
and vertex corrections. The diagrams with the intermediate 
$\Delta$ states are not shown.}
\end{center}
\end{figure}
%------------------------%------------------------%-------------------%

The diagrams responsible for the renormalization of CrTPE in the nuclear
medium are shown in Fig.~\ref{CorrTPECrMedium}. There, in analogy to what was
done for the $\pi \pi$ interaction we include $\pi$ selfenergy corrections as
well as vertex corrections.
We find that the cancellation  of the off shell part of 
the vacuum $\sigma$-exchange discussed in~\cite{Oset:2000gn} 
is also exact at finite baryonic density 
but for zero momentum transfer
only (see Appendix B). The results of the derivation for $\qvec \neq 0$ 
can be summarized as follows:

1) The expressions for the triangle vertex functions $\tilde{V}_N$ and
$\tilde{V}_{\Delta}$ of Ref.~\cite{Oset:2000gn} 
are obtained in the same way replacing
the two free pion propagators by the renormalized ones.

2) The Lindhard function entering the pion selfenergy is changed to
\begin{equation}
\mathcal{U}(k) \to \mathcal{U}(k) \cdot \left[ 1 -
\frac{1}{6}  \frac{\qvec^2}{\qvec^2 + \ds {m_{\pi}^2}/{2}} 
\right] 
\end{equation}
to account for the corrections obtained at $\qvec \neq 0$. These corrections
are negligible for small $|\qvec|$ and for $|\qvec| > m_{\pi}$ of the order
of $15\%$, hence moderate in all cases.

3) The final expression for the potential $V_{\sigma}$ is given by 
Eq.~(\ref{Vcrtpesigma}), which accounts for the $\pi \pi$ rescattering, by 
substituting $G_{\pi\pi}(t)$ by $\tilde{G}_{\pi\pi}(t)$ of Eq.~(\ref{GpipiGen})
and taking the expression for the in-medium vertex function 
$V(t)=\tilde{V}_{N}(t)+\tilde{V}_{\Delta}(t)$ where
\begin{eqnarray}
\tilde{V}_{N}(t) &=&  \frac{2 \kappa_N}{\pi^2}
\int \frac{d^3 \kvec}{(2 \pi)^3}
\frac{M_N}{E(\kvec)} \Big(\kvec^2 + \kvec\qvec\Big)
\int \limits_0^{\infty} dx
\Big[\mbox{Im} \, \tilde{D}_{\pi}(x,\kvec)\Big]
\nonumber  \\ \nonumber 
\\
&&\times
\int \limits_0^{\infty} dy
\frac{x+y+E(\kvec)-M_N}
{(x+y)(x+E(\kvec)-M_N)(y+E(\kvec)-M_N)}
\Big[\mbox{Im} \, \tilde{D}_{\pi}(y,\kvec+\qvec)\Big]
\end{eqnarray}
%In the vacuum
%$\mbox{Im} \, D_{\pi}(x,\kvec) = - \pi \delta(x^2 - \kvec^2-m_{\pi}^2)$
\begin{eqnarray}
\tilde{V}_{\Delta}(t) &=& \frac{2  \kappa_{\Delta}}{\pi^2} \frac{4}{9}
\int \frac{d^3 \kvec}{(2 \pi)^3}  \frac{M_{\Delta}}{E_{\Delta}(\kvec)}
\Big(\kvec^2 + \kvec\qvec\Big) 
\int \limits_0^{\infty} dx \Big[\mbox{Im} \, \tilde{D}_{\pi}(x,\kvec)\Big] 
\nonumber \\ \nonumber \\
&&\times
\int \limits_0^{\infty} dy
\frac{x+y+E_{\Delta}(\kvec)-M_N}
     {(x+y)(x+E_{\Delta}(\kvec)-M_N)
      (y+E_{\Delta}(\kvec)-M_N)}
\Big[\mbox{Im} \, \tilde{D}_{\pi}(y,\kvec+\qvec)\Big] 
\end{eqnarray}
The coupling constants $\kappa_{n}$ are defined by
\begin{equation}
\label{KappaConstants}
\kappa_N =\left(\frac{D+F}{2 f_{\pi}}\right)^2, ~~~~~
\kappa_{\Delta} = \left(\frac{3}{\sqrt{2}} \frac{D+F}{2 f_{\pi}} \right)^2
\end{equation}

%------------------------Figure:TriangleGraph -----------------------%
\begin{figure}[t]
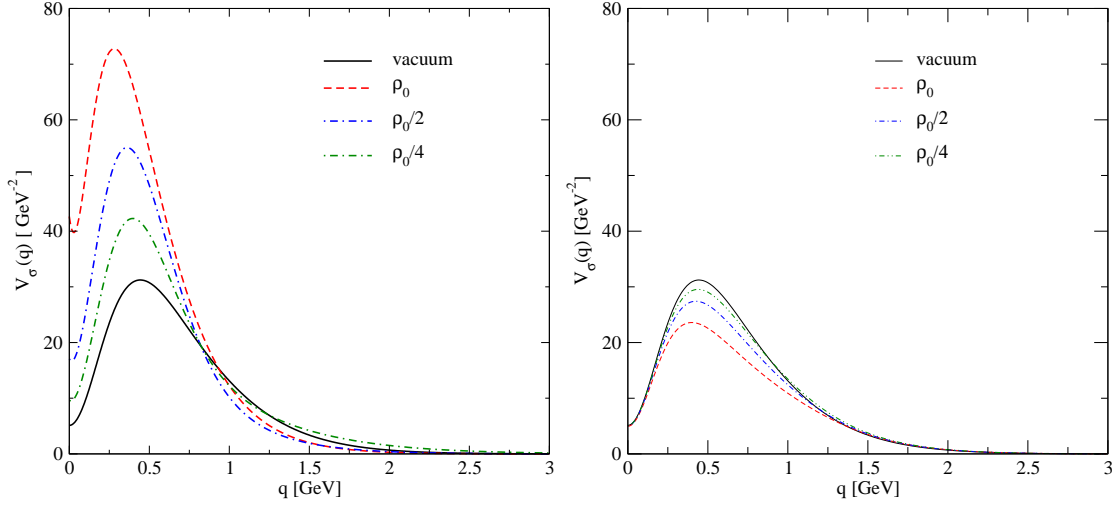

\begin{center}
\includegraphics[clip=true,width=0.45\columnwidth,angle=0.]
{Figure11a.eps}
\includegraphics[clip=true,width=0.45\columnwidth,angle=0.]
{Figure11b.eps}
\caption{\label{SigmaExchMediumMod} \footnotesize The momentum space 
unitary $\sigma$ meson 
exchange $NN$ potential 
at finite baryon density. Left panel without short range correlations. Right
panel with short range correlations.}
\end{center}
\end{figure}
%------------------------%------------------------%-------------------%

There is still one more correction to be done in order to account for short
range correlations. So far we have replaced the free pion propagator by the
renormalized one of Eq.~(\ref{InMediumGF}). However, since the interaction
between the $p-h$ bubble and the external nucleons is affected by correlations 
one should take $\mathcal{V}_l$ (see Eq.~(\ref{Vl})) instead of 
$[(D+F)^2/2f_{\pi}]^2 D_{\pi}$. Thus we would get the series
\begin{eqnarray}
D_{\pi}(k) + 
D_{\pi}(k) \, \mathcal{U}(k) \, \mathcal{V}_l(k) + 
D_{\pi}(k) \, \mathcal{U}(k) \, \mathcal{V}_l(k) \, \mathcal{U}(k)\,  
\mathcal{V}_l(k)
+ \cdots = \frac{D_{\pi}(k)}{1 - \mathcal{U}(k) \, \mathcal{V}_l(k)} \nonumber \\ = 
\frac{\tilde{D}_{\pi}(k)}
{\ds 1 - \left(\frac{D+F}{2 f_{\pi}}\right)^2 g' F^2(k) \mathcal{U}(k)}
\end{eqnarray}
where $D_{\pi}$ and $\tilde{D}_{\pi}$ are the free and dressed pion 
propagators, respectively. Hence, the expressions for $\tilde{V}_N$ and 
$\tilde{V}_{\Delta}$
get modified by including inside the $\int d^3 \kvec$ integral the 
factors
\begin{equation}
\frac{1}
{\ds 1 - \left(\frac{D+F}{2 f_{\pi}}\right)^2 g' F^2(k) \mathcal{U}(k)}
\times \frac{1}
{\ds 1 - \left(\frac{D+F}{2 f_{\pi}}\right)^2 g' F^2(k+q) \mathcal{U}(k+q)}
\end{equation}

Our results for $\sigma$-meson exchange in the momentum space are shown
if Fig.~\ref{SigmaExchMediumMod} for both cases, no short range correlations 
(left) and with short range correlations (right).

\section{Uncorrelated two pion exchange}
In this section we consider another sort of intermediate distance contributions to the
$NN$ force generated by the uncorrelated two pion exchange. The material 
presented here for the vacuum $NN$ scattering is standard and we
merely generalize it to the nuclear medium.

%------------------------Figure:TriangleGraph -----------------------%
\begin{figure}[t]
\begin{center}
\includegraphics[clip=true,width=1\columnwidth,angle=0.]
{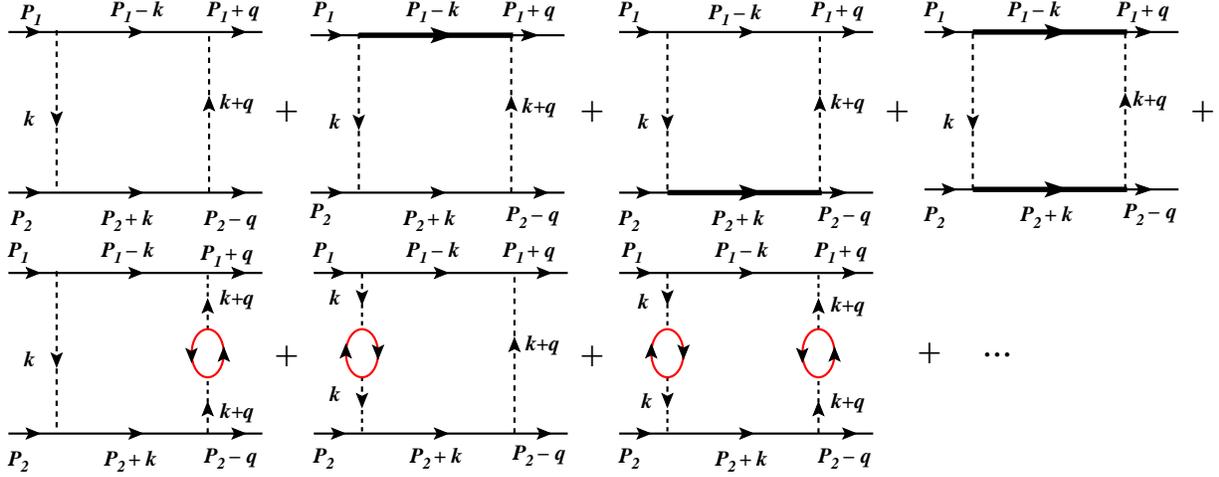}
\caption{\label{PlanarMedium} \footnotesize 
The planar box diagrams involving the
  nucleon and $\Delta$ intermediate states.}
\end{center}
\end{figure}
%------------------------%------------------------%-------------------%

\begin{figure}[t]
\begin{center}
\includegraphics[clip=true,width=1\columnwidth,angle=0.]
{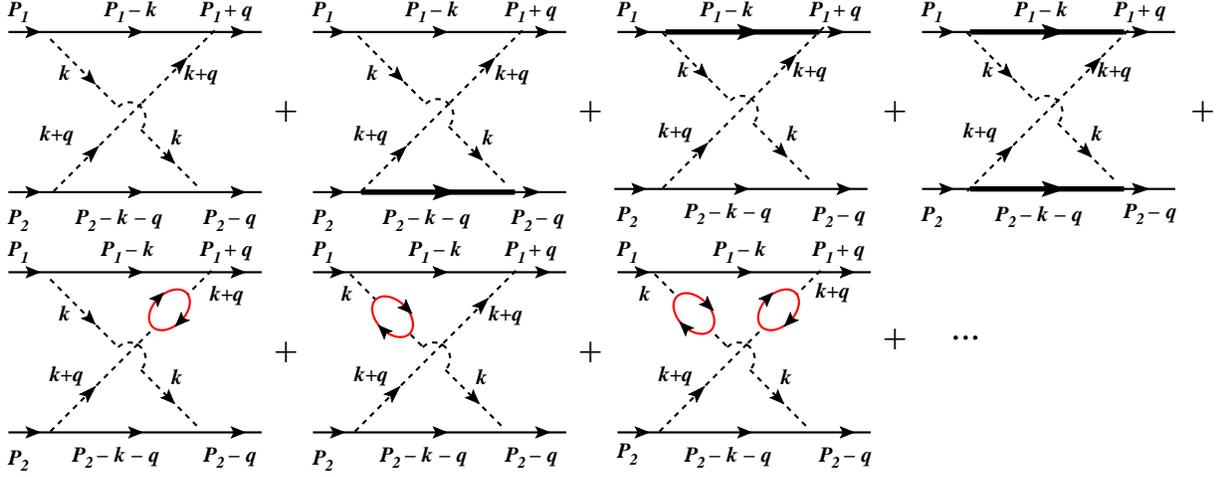}
\caption{\label{CrossedMedium} \footnotesize 
The crossed box diagrams with nucleon and $\Delta$ intermediate 
states.}
\end{center}
\end{figure}

In the perturbative expansion of the $NN$ force 
we must to take into account the planar
and crossed box diagrams shown in Figs.~\ref{PlanarMedium} 
and~\ref{CrossedMedium}, respectively. We will discuss the scalar-isoscalar
part of this contributions only. The contribution of the isovector exchange
is small and can be found, for instance, in Ref.~\cite{Jido:2001am}

In vacuum, the expression for the planar box diagrams, 
Fig.~\ref{PlanarMedium},  including the nucleon pole and $N\Delta$ and 
$\Delta \Delta$ intermediate states reads
\begin{equation}
\label{Planar}
-i V^{(P)}_{NN} =
\sum_{n,m} {\kappa_n \kappa_m }
\int \frac{d^3 \kvec}{(2 \pi)^{3}}
\Big[\Sigmacapvec^{(1)}_n \cdot (\kvec +\qvec)\Big] 
\Big[\Sigmacapvec^{(1)}_n \cdot \kvec\Big]^{\dagger}
\Big[\Sigmacapvec^{(2)}_m \cdot (\kvec +\qvec)\Big] 
\Big[\Sigmacapvec^{(2)}_m \cdot \kvec\Big]^{\dagger}
  \,
\Big\{\mathcal{R}_{nm}^{(P)}(\kvec, \qvec) \, I_{nm}^{(P)}\Big\}
\end{equation}
In Eq.~(\ref{Planar}) the sum over $n,m = N,\Delta$ is assumed and the 
spin transition operators are
$\Sigmacapvec_N = \sigmavec$, $\Sigmacapvec_{\Delta} = \Svec$.
The coupling constants $\kappa_{n}$ are defined in Eq.~(\ref{KappaConstants}).
The isospin factors are given by $I_{nm}^{(P)}$
for which we find
\begin{equation}
\label{IPL}
I_{NN}^{(P)}=3-2\tauvec_{1}\tauvec_{2},~~~
I_{N\Delta}^{(P)}=2+\frac{2}{3}\tauvec_{1}\tauvec_{2}, ~~~
I_{\Delta\Delta}^{(P)}=\frac{4}{3}-\frac{2}{9}\tauvec_{1}\tauvec_{2}
\end{equation}
The function $\mathcal{R}_{nm}^{(P)}$ in Eq.~(\ref{Planar}) contains the integration over the time-like component of the four vector $k$.
\begin{eqnarray}
\label{RnmFull}
 \mathcal{R}_{nm}^{(P)}(\kvec, \qvec) &=& %(i)^4
\int \limits_{-\infty}^{\infty}
\frac{d\, k_0}{2 \pi}
 D_{\pi}(k+q) \, D_{\pi}(k) \, S_n(P_1-k) \, S_m(P_2+k)
 \end{eqnarray}
where $D_{\pi}$ and $S_i$ is the pion and nonrelativistic baryon propagators,
respectively. $S_i$ is given by
\begin{equation}
iS_i(P) = \frac{M_i}{\sqrt{\Pcapvec^2+M_i^2}} 
\frac{i}{P_0-\sqrt{\Pcapvec^2+M_i^2}+i0^+}
\end{equation}
 This integration can be carried out explicitely
 \begin{eqnarray}
\label{RnmFull1}
 \mathcal{R}_{nm}^{(P)}(\omega_1, \omega_2) &=& \frac{i}{2}
\Big[
\Big(\omega_1^2+\omega_2^2 + 3\omega_1\omega_2 +
[E_n-E][E_m-E]\Big)(E_n+E_m-2 E)
  \nonumber\\
&&+  \Big(\omega_1 +\omega_2\Big)
\Big(2\omega_1\omega_2+[E_n+E_m-2 E]^2\Big)\Big]  \nonumber\\
&&\times
\frac{1}{\omega_1\omega_2 (\omega_1 + \omega_2)}
 \left\{\frac{1}{
E_n+E_m- 2 E - i 0^+}\right\}
   \nonumber\\
&&\times  \frac{M_n}{E_n
[\omega_1 + E_n-E][\omega_2 + E_n-E]} \nonumber\\
&&\times
\frac{M_m}{E_m
[\omega_1 + E_m-E][\omega_2 + E_m-E]
}
\end{eqnarray}
Recall that $q=(0,\qvec)$, $P_1=(E,\Pcapvec)$, $P_2=(E,-\Pcapvec)$  
in $NN$ c.m. frame and in Eq.~(\ref{RnmFull1})
\begin{equation}
\label{Kin}
\omega_1=\sqrt{\kvec^2+M_{\pi}^2}, ~~
\omega_2=\sqrt{(\kvec+\qvec)^2+M_{\pi}^2},~~
E_n=\sqrt{(\Pcapvec-\kvec)^2+M_n^2}, ~~
E_m=\sqrt{(\Pcapvec-\kvec)^2+M_m^2},
\end{equation}
are the on-shell energies of intermediate pions and baryons with c.m. energy
of the initial nucleons $E=\sqrt{\Pcapvec^2+m}$.

In the nuclear matter  $\mathcal{R} \to  \tilde{\mathcal{R}}$, and the
expression for $\tilde{\mathcal{R}}$ is obtained by using the dispersion
representation
for the in-medium pion propagator
\begin{eqnarray}
 \tilde{\mathcal{R}}_{nm}^{(P)}(\kvec, \qvec, \rho)
&=& \frac{1}{\pi^2} \int \limits_{0}^{\infty} dx^2 \int \limits_{0}^{\infty}  dy^2 \,
{\mathcal{R}}_{nm}^{(P)}(x,y) \,
\mbox{Im}\tilde{D}_{\pi}(x,\kvec,\rho) \, \mbox{Im} \tilde{D}_{\pi}(y,\kvec+\qvec,\rho)
\end{eqnarray}
where ${\mathcal{R}}_{nm}^{(P)}(x,y)$ is given by Eq.~(\ref{RnmFull1}) 
with substitution $\omega_1 \to x$ and $\omega_2 \to y$. 

%------------------------Figure:TriangleGraph -----------------------%
%------------------------%------------------------%-------------------%

For the crossed box diagrams, Fig.~\ref{CrossedMedium}, we have
\begin{equation}
\label{Crossed}
-i V^{(C)}_{NN} =
\sum_{n,m} \kappa_n \kappa_m
\int \frac{d^3 \kvec}{(2 \pi)^3}
\Big[\Sigmacapvec^{(1)}_n \cdot(\kvec +\qvec)\Big] 
\Big[\Sigmacapvec^{(1)}_n \cdot \kvec\Big]^{\dagger}
\Big[\Sigmacapvec^{(2)}_m \cdot \kvec\Big] 
\Big[\Sigmacapvec^{(2)}_m \cdot(\kvec +\qvec)\Big]^{\dagger}
 \,
\Big\{\mathcal{R}_{nm}^{(C)}(\kvec, \qvec) \, I_{nm}^{(C)}\Big\}
\end{equation}
The isospin factors are the following
\begin{equation}
\label{ICR}
I_{NN}^{(C)} = 3+2\tauvec_{1}\tauvec_{2}, ~~~
I_{N\Delta}^{(C)} = 2-\frac{2}{3}\tauvec_{1}\tauvec_{2}, ~~~
I_{\Delta\Delta}^{(C)} = \frac{4}{3}+\frac{2}{9}\tauvec_{1}\tauvec_{2}
\end{equation}
and the vacuum expression for $\mathcal{R}_{nm}^{(C)}(\kvec, \qvec)$  
is given by
\begin{eqnarray}
\label{Rcrossed}
 \mathcal{R}_{nm}^{(C)}(\kvec, \qvec) &=&
\int \limits_{-\infty}^{\infty}
\frac{d\, k_0}{2 \pi}
 D_{\pi}(k+q) \, D_{\pi}(k) \, S_n(P_1-k) \, S_m(P_2-k-q)
\end{eqnarray}
The integration can be done analytically and our result reads
\begin{eqnarray}
\mathcal{R}_{nm}^{(C)}(\omega_1, \omega_2) &=& \frac{i}{2}
\Big[\omega_1^2+\omega_2^2 + \omega_1\omega_2 +(\omega_1 + \omega_2) 
(\tilde{E}_n+E_m-2 E)+
(\tilde{E}_n-E)(E_m-E)\Big]  \nonumber\\
&&\times
\frac{1}{\omega_1\omega_2 (\omega_1 + \omega_2)}
   \nonumber\\
&&\times  \frac{M_n}{\tilde{E}_n
[\omega_1 + \tilde{E}_n-E][\omega_2 + \tilde{E}_n-E]} \nonumber\\
&&\times
\frac{M_m}{E_m
[\omega_1 + E_m-E][\omega_2 + E_m-E]
}
\end{eqnarray}
here $\omega_1,\omega_2$, $E$ and $E_m$ are defined in Eq.~(\ref{Kin}) and
\begin{equation}
\tilde{E}_n = \sqrt{(\Pcapvec+\kvec + \qvec)^2 + M_n^2}
\end{equation}
The corresponding expression for the in-medium crossed box diagrams takes 
the form
\begin{equation}
\tilde{\mathcal{R}}_{nm}^{(C)}(\kvec, \qvec, \rho) =
\frac{1}{\pi^2} \int \limits_{0}^{\infty} d x^2 \int \limits_{0}^{\infty}  
d y^2
\, \mathcal{R}_{nm}^{(C)}(x,y) \, \mbox{Im} \tilde{D}_{\pi}(x,\kvec,\rho)
\, \mbox{Im} \tilde{D}_{\pi}(y,\kvec+\qvec,\rho)
\end{equation}

%------------------------Figure:Uncorrelated Exchange -----------------------%
\begin{figure}[t]
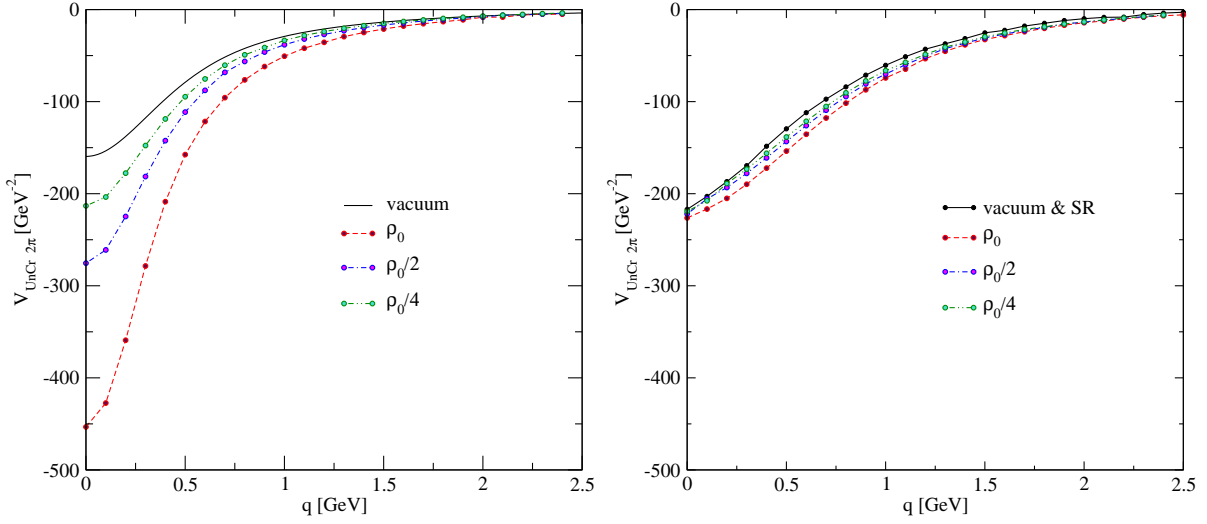

\begin{center}
\includegraphics[clip=true,width=0.485\columnwidth,angle=0.]
{Figure14a.eps}
\includegraphics[clip=true,width=0.485\columnwidth,angle=0.]
{Figure14b.eps}
\caption{\label{UncorMAINMedium} \footnotesize 
The momentum space uncorrelated two-pion exchange at finite baryonic density. 
The left panel correspond to normal pion dressing 
without SR correlations and right panel with SR correlations.}
\end{center}
\end{figure}
%------------------------%------------------------%-------------------%

For the generic case of non-vanishing initial momenta the analytical structure 
of Eq.~(\ref{RnmFull1}) is driven by the term in figure brackets
\begin{equation}
\label{NNenDenom}
\sim \frac{1}{E_n+E_m- 2 E - i 0^+}
\end{equation}
Here we have to pay a special attention to the case where two intermediate $NN$
states $n=m=N$ appear, because  in time ordered perturbation theory 
these diagrams are generated by iterations of the OPEP in
a Lippmann-Schwinger equation (LSE). 
Considering the $NN$ intermediate state only, the iterated TPE can be easily
identified and comes from the nucleon pole in Eq.~(\ref{RnmFull}) 
(in the lower half of
the complex plane) corresponding
to $k_0=E_N-E-i0^+$ with $E=\sqrt{\Pcapvec^2+M_N^2}$
\begin{equation}
\label{NucleonPole}
\mathcal{R}_{NN}^{(P)N-pole}(\omega_1,\omega_2) =
\frac{-i}{2} \left(\frac{M_N}{E_N}\right)^2 \frac{1}{E-E_N+i0^+}
\left[\frac{1}{(E-E_N)^2-\omega_1^2} \frac{1}{(E-E_N)^2-\omega_2^2} \right]
\end{equation}
Expanding Eq.~(\ref{NucleonPole}) in powers of $1/M_N$
\begin{equation}
\mathcal{R}_{NN}^{(P)N-pole}(\omega_1,\omega_2) \simeq -i 
\frac{1}{\omega_1^2\omega_2^2}
\frac{M_N}{\ds \Pcapvec^2 - (\Pcapvec-\kvec)^2 + i0^+} + \mathcal{O}(1/M_N)
\end{equation}
and inserting the leading order result in Eq.~(\ref{Planar}) one can get
the second order term in the non-relativistic Lippmann-Schwinger equation.

It is instructive to derive the contributions 
of the two remaining poles (in lower half-plane) from the pion propagators
\begin{equation}
\label{RnnIrreduc}
\mathcal{R}_{NN}^{(P)\pi-poles}(\kvec, \qvec) = \frac{-i}{2\omega_1 \omega_2}
\frac{\omega_1^2 + \omega_2^2 + \omega_1 \omega_2 - (E-E_N)^2}
{(\omega_1+\omega_2)(\omega_1^2-(E-E_N)^2)(\omega_2^2-(E-E_N)^2)} 
\frac{M^2_N}{E^2_N}
\end{equation}
After the expansion of this result in powers of $1/M_N$ we get
\begin{equation}
\label{Pipoles}
\mathcal{R}_{NN}^{(P)\pi-poles}(\kvec, \qvec) \simeq -\frac{i}{2}
\frac{\omega_1^2 + \omega_2^2 + \omega_1 \omega_2}
{\omega_1^3 \omega_2^3(\omega_1+\omega_2)} 
+ \mathcal{O}(1/M_N)
\end{equation}
In this limit Eq.~(\ref{Pipoles})  cancels
exactly the corresponding crossed box diagram with two intermediate nucleons
in the isoscalar channel.
Indeed, considering the crossed box diagram with two intermediate nucleons, 
the leading
term of $\mathcal{R}_{NN}^{(C)}(\kvec, \qvec)$
in the $1/M_N$ expansion is given by
\begin{equation}
\label{PipolesCR}
\mathcal{R}_{NN}^{(C)}(\kvec, \qvec) \simeq \frac{i}{2}
\frac{\omega_1^2 + \omega_2^2 + \omega_1 \omega_2}
{\omega_1^3 \omega_2^3(\omega_1+\omega_2)} + \mathcal{O}(1/M_N)
\end{equation}
Note that for the isovector $\pi \pi$ exchange because of the different sign 
of the $\tauvec_1 \tauvec_2$ term in $I_{NN}^{(P)}$, $I_{NN}^{(C)}$ in
Eqs.~(\ref{IPL}) and~(\ref{ICR}) these two contributions would add.
The cancellation discussed above hold also in the nuclear medium.

The result of the vacuum scalar-isoscalar $NN$ force generated
by the planar and crossed box diagrams, with the nucleon pole diagrams excluded,
is shown in the left panel of Fig.~\ref{UncorMAINMedium}  by the 
solid curve. It is in agreement with Ref.~\cite{Jido:2001am} where 
it was shown that the consistent use of the cut off regularization
in both the correlated two pion exchange, and uncorrelated two pion exchange, 
together with the contribution of a repulsive $\omega$-exchange
lead to a scalar-isoscalar potential in good agreement with the 
Argonne~\cite{Wiringa:1994wb}
potential in the whole range of relevant distances.
The corresponding results for the nuclear matter are shown in 
Fig.~\ref{UncorMAINMedium} (left) for three densities 
$\rho_0$, $\rho_0/2$ and $\rho_0/4$. 
Qualitatively the behavior is similar to the 
vacuum case but with a strong enhancement toward the small momentum transfer.
As we have seen this feature is generic in present calculations. 
Uncorrelated $\pi \pi$ exchange becomes extremely attractive at intermediate
distances.

The spin sum over intermediate baryon states of the 
$\sim \Sigmacapvec \cdot \kvec$ operator in Eqs.~(\ref{Planar}) 
and~(\ref{Crossed})
gives in the scalar channel
\begin{equation}
\beta \left[ (\kvec + \qvec) \cdot \kvec \right]^2 
\end{equation}
where $\beta=1$ $(NN)$, $\beta=2/3$ $(N\Delta)$ and $\beta=4/9$ 
$(\Delta\Delta)$. Now we again wish to take into account the short range 
correlations
\begin{eqnarray}
\left(\frac{D+F}{2 f_{\pi}}\right)^4 F^2(\kvec)F^2(\kvec+\qvec)
{D}_{\pi}(\kvec+\qvec){D}_{\pi}(\kvec) 
\left[ (\kvec + \qvec) \cdot \kvec \right]^2  \nonumber \\
\Longrightarrow \tilde{\mathcal{W}}_l(k+q) \tilde{\mathcal{W}}_l(k)
\left[ (\widehat{\kvec + \qvec}) \cdot \hat{\kvec} \right]^2  \nonumber \\
+ \left[  \tilde{\mathcal{W}}_l(k+q)\tilde{\mathcal{W}}_t(k)
 + 
\tilde{\mathcal{W}}_l(k) \tilde{\mathcal{W}}_t(k+q)\right]
\left\{  1 - \left[ (\widehat{\kvec +\qvec}) 
\cdot \hat{\kvec}\right]^2 \right\}
\nonumber \\
+ \tilde{\mathcal{W}}_t(k+q) \tilde{\mathcal{W}}_t(k)
\left\{ 1 + \left[ (\widehat{\kvec +\qvec})
\cdot \hat{\kvec} \right]^2 \right\} \nonumber \\
\end{eqnarray}
where
\begin{equation}
\tilde{\mathcal{W}}_i(k) = \frac{\mathcal{V}_i(k)}
{1-\mathcal{U}(k)\mathcal{V}_i(k) }
\end{equation}
where $\mathcal{U}(k)$ is the Lindhard function and 
$\mathcal{V}_l$ and $\mathcal{V}_t$ are given by Eq.~(\ref{Vl}) 
and~~(\ref{Vt}), respectively.

%\section{Iterated two pion exchange at finite density}

\section{Results and discussions}
By comparing the results in Fig.~\ref{OPEPMedium}  and 
Fig.~\ref{OPEPMediumModif} we observe that the effect of
correlations has been essential and reduces drastically the medium effects
found in Fig.~\ref{OPEPMedium} without corrections. We observe that the medium 
corrections
weaken the strength of the central part of the OPEP. On the other hand 
the effect of the medium corrections on the tensor part are more moderate.

  The $\sigma$ exchange presents similar features as we can see in 
  Fig.~\ref{SigmaExchMediumMod}.
  There we see that the medium effects in the absence of short range
  correlations are rather large and increase the strength of the interaction in
  about a factor two at $\rho=\rho_0$.  However, as soon as the correlations
  are taken into account the medium modifications are reduced drastically 
  and at $\rho=\rho_0$ just reduce the strength in about 25 percent.

  The medium effects in the uncorrelated two pion exchange are also
  relatively small, with respect to the size of the interaction 
  in the vacuum, see Fig. 14 (right).  Once again
  the consideration of the correlations has been essential and reducing the
  size of the medium effects.  However, the
  fact that the strength of this interaction is bigger than that of the
  correlated two pion exchange makes the absolute correction relevant and of
  the same size as the corrections discussed before.

%------------------------Figure:Uncorrelated Exchange -----------------------%
\begin{figure}[t]
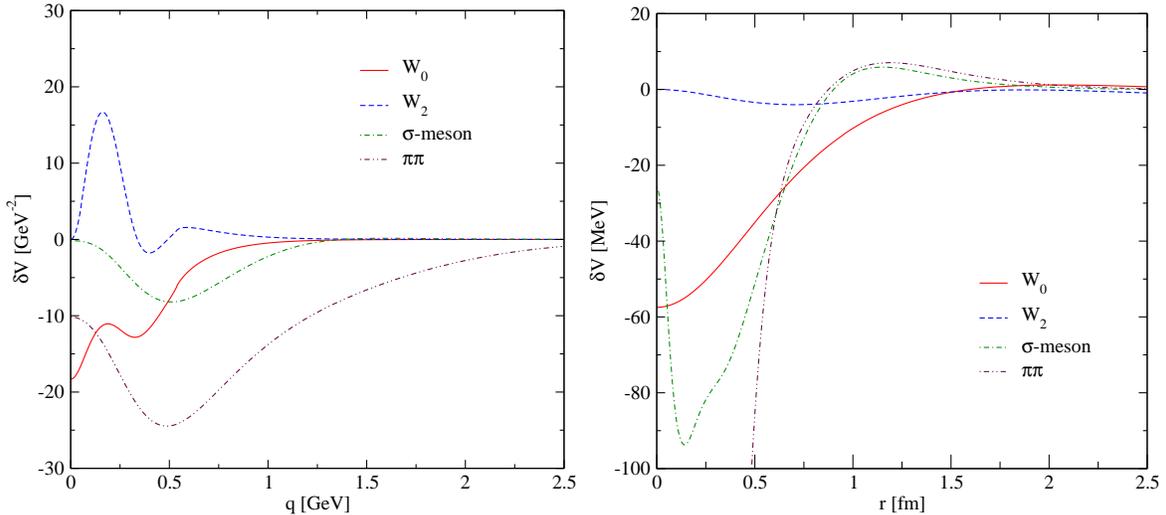

\begin{center}
\includegraphics[clip=true,width=0.47\columnwidth,angle=0.]
{Figure15a.eps}
\includegraphics[clip=true,width=0.47\columnwidth,angle=0.]
{Figure15b.eps}
\caption{\label{UncorMAINmediumMM} \footnotesize 
The modification of the $NN$ interaction at normal nuclear matter density
in momentum space (left panel) and in configuration space (right panel).
Shown are the central part of the OPEP (solid curve), tensor part of the OPEP
(dashed curve), correlated $\sigma$-meson exchange (dot-dashed curve)
and uncorrelated $\pi \pi$ exchange (dot-dot-dashed curve).}
\end{center}
\end{figure}
%------------------------%------------------------%-------------------%

  Since we have taken also the effect of correlations in the free part of the
 interaction and any realistic calculation of the binding energy of matter will
 explicitly account for these correlations, the use of our modified in-medium
 potentials would lead to double-counting if any of these methods is used.  For
 this purpose the relevant results from this work should be the differences
 between the in-medium potential and the free one. In this case we start 
 already
 with one bubble and the correlations account for the repulsion between  the
 external leg and the one in the bubble.  The two external legs still have to
 be correlated and this will be done with the use, for instance, of the 
 ordinary Bethe Golstone equation.

     After this discussion of the medium effects in the different terms, we
 show the differences in momentum space between the medium and free parts of
 the different terms in Fig.~\ref{UncorMAINmediumMM} (left). 
 The results are shown at $\rho=\rho_0$
 and we see that these corrections are moderate, but they could have a relevant
 role in the binding of matter. In order to have a qualitative idea of the
 relevance of these corrections to the potentials we rewrite them in coordinate
 space and show these results in Fig.~\ref{UncorMAINmediumMM} (right).  
 The effects seem sizeable at
 short distances, but this will be irrelevant in any realistic calculation of
 binding energies since the consideration of the short range correlations
 between the external legs will make this interaction inoperative.  More
 interesting is the strength of the corrections around 1 fm,  and there we see
 that all them are relatively small, of the order of 20 MeV or less, the
 biggest one being the central part of the OPE.  However, given the size of the
 empirical scalar isoscalar attraction \cite{Wiringa:1994wb} which
 is of the order of 20 MeV at intermediate densities, the corrections found 
 here are not negligible. It would be thus interesting to see the effects of 
 the results obtained here in observables like the binding of nuclear matter 
 and other properties, which we hope to stimulate with the results obtained 
 here.

\section{Conclusions}
   In this paper we studied the modification of the one pion exchange, as well
as two pion exchange  potential inside a nuclear medium.   For this purpose we
separated the two pion exchange into an uncorrelated two pion exchange and the
correlated two pion exchange.   We study both in the scalar isoscalar channel,
which is by large the most important one generated by this interaction. The
correlated two pion exchange gave rise to the equivalent of the $\sigma$
exchange in other models and we studied the medium modification  to it.  On
the
other hand, for the uncorrelated two pion exchange we have followed a
traditional approach in which only terms with at least one intermediate
$\Delta$
state are considered.

    One of the important findings here was the effect played by the NN short
 range correlations which drastically moderated the medium corrections to the
 potential.  In the absence of these, the corrections where unrealistically
 large.  Yet, even if relatively moderate, the medium corrections found in this
 paper are sizeable enough to have relevant repercussions in the binding and
 other properties of nuclear matter and we would like to encourage calculations
 in  this direction.

\section*{Acknowledgments}
We would like to acknowledge A.~Ramos for a critical reading of the manuscript
and useful suggestions.
This work is partly supported by DGICYT contract number BFM2003-00856,
and the E.U. EURIDICE network contract no. HPRN-CT-2002-00311.
This research is part of the EU Integrated Infrastructure Initiative
Hadron Physics Project under contract number RII3-CT-2004-506078.

\begin{appendix}
\section{Feynman rules for vertices}
\begin{figure}[h]
\begin{center}
\includegraphics[clip=true,width=0.7\columnwidth,angle=0.]{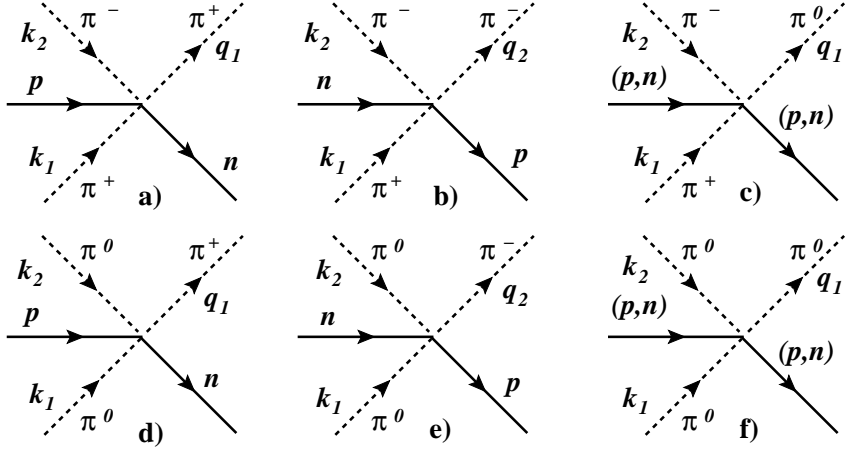}
\caption{\label{3PiNNVertCharge} \footnotesize
         Contact terms appearing in the construction of the 
         $\pi N \to \pi \pi N$ transition amplitude.}
\end{center}
\end{figure}

The set of Feynman diagrams shown in Fig.~\ref{3PiNNVertCharge}
appear in the construction of the $\pi N \to \pi \pi N$ transition amplitude.
The corresponding $3\pi NN$ vertex functions are given by
%-------------------Figure: Verteces in the charged basis-------------

%------------------------Figure: Sigma PiPi--------------------------

\begin{eqnarray}
i L_a &=& \frac{D+F}{2}  \frac{2 \sqrt{2}}{12 f_{\pi}^3}
\Big[ \sigmavec (\qvec_1 + 2 \kvec_1 - \kvec_2) \Big] \\
i L_b &=& \frac{D+F}{2}  \frac{2 \sqrt{2}}{12 f_{\pi}^3}
\Big[ \sigmavec (\qvec_2 + 2 \kvec_2 - \kvec_1) \Big] \\
i L_c &=& \pm \frac{D+F}{2}  \frac{2}{12 f_{\pi}^3}
\Big[ \sigmavec (2 \qvec_1 + \kvec_1 + \kvec_2) \Big] \\
i L_d &=& \frac{D+F}{2}  \frac{2 \sqrt{2}}{12 f_{\pi}^3}
\Big[ \sigmavec (2\qvec_1 + \kvec_1 + \kvec_2) \Big] \\
i L_e &=& \frac{D+F}{2}  \frac{2 \sqrt{2}}{12 f_{\pi}^3}
\Big[ \sigmavec (2\qvec_2 + \kvec_1 + \kvec_2) \Big] \\
i L_f &=& 0
\end{eqnarray}

\section{Correlated exchange in the medium}
In this Appendix we demonstrate the cancellation of the off shell part of the
correlated two pion exchange in the scalar isoscalar channel. For simplicity
we consider the nucleon intermediate states only. The generalization
to $N\Delta$ and $\Delta \Delta$ is straightforward. The 
diagrams are shown in Fig.~\ref{CorrTPECrMedium} and corresponding amplitudes
are given by
\begin{eqnarray}
T_a &=& T_a^{on} + T_a^{(1)} + T_a^{(2)} \\
-i T_a^{on}
&=& 6  \left(\frac{D+F}{2 f_{\pi}}\right)^4
 V^{on}_{\pi}(t)  \Big[V_{N\pi\pi}(t) \Big] \nonumber\\
&&\times
 \int \frac{d^4 k}{(2 \pi)^4}  \Big[\sigmavec (\kvec+\qvec) \Big]
 \Big[\sigmavec \kvec \Big] 
 \Big[\kvec^2  \mathcal{U}(k)\Big] S_F(P_1-k) D^2(k) D(k+q) \\
-i T_a^{(1)} &=& 2 \left(\frac{D+F}{2 f_{\pi}}\right)^4
 \frac{1}{f_{\pi}^2}  \Big[V_{N\pi\pi}(t) \Big] \nonumber\\
&&\times
  \int \frac{d^4 k}{(2 \pi)^4}
 \Big[\sigmavec (\kvec+\qvec) \Big]
 \Big[\sigmavec \kvec \Big] 
 \Big[\kvec^2  \mathcal{U}(k)\Big] S_F(P_1-k) D(k) D(k+q)  \\
-i T_a^{(2)} &=& 2 \left(\frac{D+F}{2 f_{\pi}}\right)^4
 \frac{1}{f_{\pi}^2}
  \Big[V_{N\pi\pi}(t) \Big] \nonumber\\
&&\times \int \frac{d^4 k}{(2 \pi)^4}
\Big[\sigmavec (\kvec+\qvec) \Big]
 \Big[\sigmavec \kvec \Big]
 \Big[\kvec^2  \mathcal{U}(k)\Big] S_F(P_1-k) D^2(k) \\ \nonumber
\end{eqnarray}
\begin{eqnarray}
T_b &=& T_b^{on} + T_b^{(1)} + T_b^{(2)} \\
-i T_b^{on} &=& 6  \left(\frac{D+F}{2 f_{\pi}}\right)^4
 V_{\pi}^{on}(t) \Big[V_{N\pi\pi}(t) \Big] \nonumber\\
&&\times
 \int \frac{d^4 k}{(2 \pi)^4}  \Big[\sigmavec (\kvec+\qvec) \Big]
 \Big[\sigmavec \kvec \Big] 
 \Big[(\kvec + \qvec)^2  \mathcal{U}(k+q)\Big] S_F(P_1-k) D(k) D^2(k+q)
 \\
-iT_b^{(1)} &=& 2 \left(\frac{D+F}{2 f_{\pi}}\right)^4
 \frac{1}{f_{\pi}^2}  \Big[V_{N\pi\pi}(t) \Big] \nonumber\\
&&\times
 \int \frac{d^4 k}{(2 \pi)^4}  \Big[\sigmavec (\kvec+\qvec) \Big]
 \Big[\sigmavec \kvec \Big] 
 \Big[(\kvec + \qvec)^2  \mathcal{U}(k+q)\Big] S_F(P_1-k) D(k) D(k+q)
\\
-iT_b^{(2)} &=& 2 \left(\frac{D+F}{2 f_{\pi}}\right)^4
 \frac{1}{f_{\pi}^2} \Big[V_{N\pi\pi}(t) \Big] \nonumber\\
&&\times
 \int \frac{d^4 k}{(2 \pi)^4}  \Big[\sigmavec (\kvec+\qvec) \Big]
 \Big[\sigmavec \kvec \Big]
 \Big[(\kvec + \qvec)^2  \mathcal{U}(k+q)\Big] S_F(P_1-k) D^2(k+q)
\nonumber \\
-i T_c &=& -i T_a(k \to l, q \to - q, P_1 \to P_2)
\nonumber \\
-i T_d &=& -i T_b(k \to l, q \to - q, P_1 \to P_2)
\end{eqnarray}
where $V_{N\pi\pi}(t)$ is the triangle loop integral
\begin{equation}
V_{N\pi\pi}(t) = i \left(\frac{D+F}{2 f_{\pi}}\right)^2
 \int \frac{d^4 l}{(2 \pi)^4}  \Big[\sigmavec (\lvec-\qvec) \Big]
 \Big[\sigmavec \lvec \Big]
 S_F(P_2-l) D(l-q) D(l)
\end{equation}
and the on-mass-shell $\pi \pi$ scattering amplitude is
\begin{equation}
V^{on}_{\pi}(t) = - \frac{1}{f_{\pi}^2}\Big(\ds t -  \frac{M_{\pi}^2}{2}\Big)
\end{equation}

\begin{eqnarray}
-i T_e &=& - \left(\frac{D+F}{2 f_{\pi}}\right)^4
 \frac{1}{f_{\pi}^2}  \Big[V_{N\pi\pi}(t) \Big] \nonumber\\
&&\times
 \int \frac{d^4 k}{(2 \pi)^4}  \Big[\sigmavec (\kvec+\qvec) \Big]
 \Big[\sigmavec \kvec \Big] 
 \Big[( 3 \kvec \qvec  + 2 \kvec^2)  
\mathcal{U}(k)\Big] S_F(P_1-k) D(k) D(k+q) \\
%\end{equation}
%\begin{equation}
-i T_f &=& \left(\frac{D+F}{2 f_{\pi}}\right)^4
 \frac{1}{f_{\pi}^2}  \Big[V_{N\pi\pi}(t) \Big] \\
&&\times
 \int \frac{d^4 k}{(2 \pi)^4}  \Big[\sigmavec (\kvec+\qvec) \Big]
 \Big[\sigmavec \kvec \Big]
 \Big[( \qvec^2 - 2 \kvec^2 - \kvec \qvec ) 
\mathcal{U}(k+q)\Big] S_F(P_1-k) D(k)
 D(k+q) \nonumber \\
-i T_g &=& -i T_e(k \to l, q \to - q, P_1 \to P_2) \\
-i T_h &=& -i T_f(k \to l, q \to - q, P_1 \to P_2) \\
%\end{equation}
%\begin{equation}
-i T_i &=&  \left(\frac{D+F}{2 f_{\pi}}\right)^4
 \frac{1}{f_{\pi}^2}  \Big[V_{N\pi\pi}(t) \Big]
 \int \frac{d^4 k}{(2 \pi)^4}  \Big[\sigmavec (\qvec-2\kvec) \Big]
\Big[\sigmavec \kvec \Big] 
 \Big[\kvec^2 \mathcal{U}(k)\Big] S_F(P_1-k) D^2(k) \\
-i T_j &=& - \left(\frac{D+F}{2 f_{\pi}}\right)^4
 \frac{1}{f_{\pi}^2}  \Big[V_{N\pi\pi}(t) \Big] \nonumber\\
&&\times
 \int \frac{d^4 k}{(2 \pi)^4}  \Big[\sigmavec \kvec \Big] 
\Big[\sigmavec (\qvec+2\kvec) \Big] 
 \Big[\kvec^2 \mathcal{U}(k)\Big] S_F(P_1+q-k) D^2(k) \\
-i T_k &=& -i T_i(k \to l, q \to - q, P_1 \to P_2) \\
-i T_l &=& -i T_j(k \to l, q \to - q, P_1 \to P_2) 
\end{eqnarray}
%\subsection{Cancellations}
In the present case $q=(0,\qvec)$ and for $\qvec \to 0$ the amplitude
$T_a^{(2)}$ cancels exactly $T_i$. The same cancelation is found for
$T_b^{(2)}$ and $T_j$.
The sum of $T_e$ and $T_a^{(1)}$ is given by
\begin{eqnarray}
\label{mm}
 -i T_a^{(1)} - i T_e &=& - 3 \left(\frac{D+F}{2 f_{\pi}}\right)^4
 \frac{1}{f_{\pi}^2}  \Big[V_{N\pi\pi}(t) \Big] \nonumber\\
&&\times
 \int \frac{d^4 k}{(2 \pi)^4}  \Big[\sigmavec (\kvec+\qvec) \Big]
 \Big[\sigmavec \kvec \Big]
 \Big[\kvec \qvec  \,  \mathcal{U}(k)\Big] S_F(P_1-k) D(k) D(k+q)
\end{eqnarray}
The sum of $T_f$ and $T_b^{(1)}$ takes the form
\begin{eqnarray}
\label{mn}
-i T_b^{(1)} - i T_f &=& 3 \left(\frac{D+F}{2 f_{\pi}}\right)^4
 \frac{1}{f_{\pi}^2}  \Big[V_{N\pi\pi}(t) \Big] \\
&&\times
 \int \frac{d^4 k}{(2 \pi)^4}  \Big[\sigmavec (\kvec+\qvec) \Big]
 \Big[\sigmavec \kvec \Big]
 \Big[(\kvec+\qvec) \qvec
\mathcal{U}(k+q)\Big] S_F(P_1-k) D(k) D(k+q) \nonumber
\end{eqnarray}
From this the sum of four diagrams is given by
\begin{eqnarray}
&& -i T_a^{(1)} - i T_e -i T_b^{(1)} - i T_f =
3 \left(\frac{D+F}{2 f_{\pi}}\right)^4
 \frac{1}{f_{\pi}^2}  \Big[V_{N\pi\pi}(t) \Big] \\
&&\times
 \int \frac{d^4 k}{(2 \pi)^4}  \Big[\sigmavec (\kvec+\qvec) \Big]
 \Big[\sigmavec \kvec \Big] 
 \Big[ \qvec^2 \mathcal{U}(k+q) 
+ \kvec \qvec  \Big\{\mathcal{U}(k+q)-\mathcal{U}(k)\Big\}
\Big] S_F(P_1-k) D(k) D(k+q) \nonumber
\end{eqnarray}
Alternatively, the spin flip parts of Eqs.~(\ref{mm}) and~(\ref{mn})  
can be canceled if
we change in Eq.~(\ref{mn})  
$\kvec \to -(\kvec + \qvec)$ and using again the fact that
$q=(0,\qvec)$ and  the properties of the Lindhard function 
$\mathcal{U}(p_0,\pvec)=\mathcal{U}(p_0,-\pvec)$ we get
\begin{eqnarray}
-i T_b^{(1)} - i T_f &=&  - 3 \left(\frac{D+F}{2 f_{\pi}}\right)^4
 \frac{1}{f_{\pi}^2}  \Big[V_{N\pi\pi}(t) \Big] \\
&&\times
 \int \frac{d^4 k}{(2 \pi)^4}  \Big[\sigmavec \kvec \Big]
 \Big[\sigmavec  (\kvec+\qvec)\Big]
 \Big[ \kvec \qvec
\mathcal{U}(k)\Big] S_F(P_1-k) D(k) D(k+q) \nonumber
\end{eqnarray}
and
\begin{eqnarray}
-i T_a^{(1)} - i T_e -i T_b^{(1)} - i T_f &=&
-6 \left(\frac{D+F}{2 f_{\pi}}\right)^4
 \frac{1}{f_{\pi}^2}  \Big[V_{N\pi\pi}(t) \Big] \\
&&\times
 \int \frac{d^4 k}{(2 \pi)^4}  \Big[(\kvec^2 +\kvec\qvec)\kvec \qvec\Big]
\mathcal{U}(k) S_F(P_1-k) D(k) D(k+q) \nonumber
\end{eqnarray}
Finally
\begin{eqnarray}
-i T_c^{(1)} - i T_g -i T_d^{(1)} - i T_h &=&
-6 \left(\frac{D+F}{2 f_{\pi}}\right)^4
 \frac{1}{f_{\pi}^2}  \Big[V_{N\pi\pi}(t) \Big] \\
&&\times
 \int \frac{d^4 l}{(2 \pi)^4}  \Big[(\lvec\qvec-\lvec^2)\lvec \qvec\Big]
\mathcal{U}(l) S_F(P_2-l) D(l) D(l-q) \nonumber
\end{eqnarray}
As one can see in the limit $\qvec \to 0$ we get an exact cancellation of 
the off mass shell part of the $\pi \pi$ amplitude and only $T_a^{on}$ and
$T_b^{on}$ are left.

\end{appendix}

\end{document}